\documentclass[%
 reprint,
groupedaddress,
nofootinbib,
nobibnotes,
 amsmath,amssymb,
aps,
] {revtex4-1}

\usepackage{graphicx}
\graphicspath{ {images/} }
\usepackage{dcolumn}
\usepackage{bm}
\usepackage{graphicx}
\usepackage{epsfig}
\usepackage{float}
\usepackage{array}
\usepackage{amsmath}
\usepackage{amsfonts}
\usepackage{amssymb}

\usepackage[utf8]{inputenc}
\usepackage[english]{babel}

\makeatletter
     \renewcommand\@make@capt@title[2]{%
      \@ifx@empty\float@link{\@firstofone}{\expandafter\href\expandafter{\float@link}}%
       {\textbf{#1}}\@caption@fignum@sep#2\quad}%
     \makeatother
 
\makeatletter 
\renewcommand{\fnum@figure}{\textbf{Figure~\thefigure}}
\makeatother

\usepackage{calc}
\usepackage[usenames, dvipsnames]{color}
\usepackage{tabularx}
\usepackage[colorlinks=true,citecolor=ForestGreen,linkcolor=red]{hyperref}
 
\newcolumntype{P}[1]{>{\centering\arraybackslash}p{#1}}
\newcolumntype{M}[1]{>{\centering\arraybackslash}m{#1}}

\usepackage[document]{ragged2e}
\usepackage[]{natbib}

\begin{document}
\bibliographystyle{unsrtnat}


\title{Closed-form solutions for the L\'evy-stable distribution}

\author{Karina Arias-Calluari$^{[1]}$}
\altaffiliation[]{School of Civil Engineering}
\email{kari0293@uni.sydney.edu.au}

\author{Fernando Alonso-Marroquin $^{[1,2]}$}%

\author{Michael S. Harr\'e $^{[3]}$}


\affiliation{$[1]$School of Civil Engineering, The University of Sydney, Sydney NSW 2006}%
\affiliation{$[2]$Computational Physics IfB, ETH Zurich, CH-8093, Zurich, Switzerland}
\affiliation{$[3]$Complex Systems Research Group, The University of Sydney, Sydney NSW 2006}

\date{\today}

\begin{abstract}
\justifying
The \textcolor{black}{L\'evy-stable} distribution is the attractor of distributions \textcolor{black}{which hold} power laws with infinite variance. \textcolor{black}{This distribution has been used in a variety of research areas, for example in economics it is used to model financial market fluctuations and in statistical mechanics as a numerical solution of fractional kinetic equations of anomalous transport. This function} does not have an explicit expression and no uniform solution has been proposed yet. This paper presents a uniform analytical approximation for the L\'evy-stable distribution based on matching power series expansions. For this solution, the trans-stable function is defined as an auxiliary function which removes the numerical issues of the calculations of the L\'evy-stable.
 \textcolor{black}{Then, the uniform solution is proposed as a result of an asymptotic matching between two types of approximations  called ``the inner solution" and ``the outer solution"}. Finally, the results of analytical approximation \textcolor{black}{are} compared to the numerical results of the L\'evy-stable distribution function, making this uniform solution valid to be applied as an analytical approximation.
\end{abstract}

\pacs{Valid PACS appear here}
\maketitle


\section{\label{sec:level1}Introduction}

\justifying

A wide range of natural and social phenomena exhibit  a power-law in the probability distribution of large events. These tails are characterized by the asymptotic relation $  f(x)\sim 1/x^{1+\alpha }$, where $x$ is the size of the events \textcolor{black}{\citep{turcotte1999self,malamud2004tails}. For $0<\alpha \leq  1$, the distribution has an indefinite mean value. On the other hand, for $1<\alpha\leq2$, the distribution has a defined mean value but still exhibits infinite variance \citep{Samo1994}.
These heavy-tailed distributions have been observed in economics and statistical mechanics.}

\textcolor{black}{In the field of economics, the statistics of price returns, trade size, and share volumes have been investigated. Heavy tailed distributions have been observed in the correlations of the absolute value of the S\&P 500 returns \citep{gopikrishnan2000scaling,biham2001long}, the effects of networks on price returns \citep{nobi2014effects}, daily returns of the Dow Jones index \citep{borak2005stable}, Brent crude oil returns \citep{yuan2014stable}  and the aggregate output growth rate distribution \citep{lera2018gross}. Even after applying five different estimation techniques, power law tails with  the characteristic index $\alpha$ were found on the cumulative distribution of trade size and share volumes of 252 US stocks  over the 42-year period from 1963-2005  \citep{plerou2007tests,plerou2009reply}. \textcolor{black}{To capture heavy tails} different models have been proposed to simulate the stock price dynamics. For instance, models of anomalous diffusion of option pricing were introduced as an extension of the well-known Brownian model \citep{magdziarz2011option,yura2014financial}. \textcolor{black}{These models are focused on different aspects such as capturing the dynamic of the price with waiting times (periods of stagnation) which are L\'evy-stable distributed \citep{magdziarz2011option} or on the effect of ``particles" representing agent's interaction \citep{yura2014financial}. 
The continuous counterpart of these discrete models is the Fokker-Planck equation (FPE) that is presented in terms of  fractional derivatives.} The solution of FPE gives the time evolution of the probability density function (pdf) of price return \citep{magdziarz2011option,friedrich2000quantify,yura2014financial}}.\par
 
\textcolor{black}{In the field of statistical mechanics, the diffusion equation (DE) is a fundamental equation of transport dynamics used \textcolor{black}{to describe a particle motion resulting from the interaction with a thermal heat bath \citep{metzler1999anomalous,metzler2000random,metzler2004restaurant}.} The DE defines the  probability of a particle to be at a certain position at a specific time, and its pdf is given by the Gaussian distribution \citep{jain2017levy,chaves1998fractional,tsallis1995statistical}. On the other hand, for the generalization of anomalous transport, a fractional diffusion equation (FDE) is used to describe a continuous time random walk (CTRW) model \citep{metzler2000random,janakiraman2012path}. This model generalizes the Brownian diffusion motion based on two parameters accounting for the jump length and the waiting time between two successive jumps. A long-tailed waiting time pdf ---long rests---, produces a ``subdiffusion process" \citep{metzler2000random,metzler2004restaurant}. On the opposite case, the L\'evy-stable distribution for the jump length pdf ---long jumps--- produces a ``superdiffusion process" \citep{metzler2000random,kaminska2017levy}. The anomalous diffusion under an external velocity field or a microscopic advection is studied by fractional diffusion-advection equations (FDAE) \citep{metzler2000random,metzler2004restaurant,meerschaert2004finite}. \textcolor{black}{Aditionally, the fractional Fokker-Planck equation (FFPE)} is used to study the anomalous diffusion under the influence of  an external field: electrical bias field \citep{metzler2000random,metzler2004restaurant}, periodic potentials \citep{srokowski2009fractional,nezhadhaghighi2017scaling} or a harmonic potential \citep{janakiraman2012path,jain2017levy,dybiec2017underdamped}.
The FFPE can be derived either from the generalized FDE of
continuous time random walk models or from a Langevin equation with Levy-stable noise or gaussian noise and long rests \citep{kessler2017stochastic,jain2017levy,nezhadhaghighi2017scaling}}.\par
\textcolor{black}{The previous fractional kinetic equations (FDE, FDAE and FFPE) can be solved in terms of L\'evy-stable distribution function} 
\textcolor{black}{that has an analytical solution for only two cases ---the normal and Cauchy distributions---\citep{yanovsky2000levy}}. In the remaining cases there is not a closed-form expression. \textcolor{black}{Typically the numerical solution} of the \textcolor{black}{L\'evy-stable} distribution has numerical oscillations  in \textcolor{black}{the tail of the distribution}. For some cases it displays apparent discontinuity in logarithmic plots because of negative values obtained from the numerical solution \citep{fernandez1998bayesian}. \textcolor{black}{\textcolor{black}{Consequently} numerical solutions of the} \textcolor{black}{L\'evy-stable} distribution \textcolor{black}{are} not reliable \textcolor{black}{as} the probability density function must be positive.

 Analytical expressions in terms of power series \textcolor{black}{have been} presented by different authors. Feller \citep{Feller1950}, Elliot \citep{montroll1984levy} and Zolotarev \citep{zolotarev1986one} used power series to obtain converging algorithms of the \textcolor{black}{L\'evy-stable} distribution function in two ranges, the \textcolor{black}{first for} $\alpha<1$ and the second for $\alpha>1$ for symmetric distributions. However, some of the proposed series do not converge to the \textcolor{black}{L\'evy-stable} function, and some of them are only applicable for extreme values $x\rightarrow 0$ or $x\rightarrow \infty$. 
Mantegna \citep{mantegna1994fast} presented a \textcolor{black}{similar} solution \textcolor{black}{that of} Elliot \citep{montroll1984levy} but \textcolor{black}{the}  algorithm is only valid when $x\rightarrow \infty$ and $0.75<\alpha\leq1.95$. Nolan \citep{nolan2003stable} presented  \textcolor{black}{an} algorithm \textcolor{black}{for} asymmetric distributions \textcolor{black}{of} large events $x\rightarrow \infty$ focusing \textcolor{black}{only on the} tail behaviour \textcolor{black}{of the distribution}. Thus, the \textcolor{black}{L\'evy-stable} distribution function does not have an explicit expression \citep{William1950,Johnson2003} and no uniform solution of the \textcolor{black}{L\'evy-stable} distribution has been proposed until now \citep{,nolan2003stable,zolotarev1986one,Feller1950}.\par

Due to the absence of an explicit expression, numerical solutions were developed to evaluate the \textcolor{black}{L\'evy-stable} distribution function by using numerical recursive quadrature methods  \citep{nolan2001maximum,nolan2013multivariate,liang2013survey}. \citet{nolan2001maximum,nolan2003modeling} develops a numerical solution for the estimation of \textcolor{black}{L\'evy-stable} parameters through a maximum likelihood method for each data set of $x$. However, Nolan's method converges only for $\alpha>0.4$ and the convergence to the \textcolor{black}{L\'evy-stable} distribution function seems to be not accurate enough. Despite this fact, Nolan's method constitutes an important method that is still being used \citep{nolan2013multivariate}. \par

Apart from the numerical issues in the evaluation of the \textcolor{black}{L\'evy-stable} distribution, some authors have pointed out its infinite variance as a drawback \citep{mantegna1994stochastic,koponen1995analytic,vasconcelos2004guided}. To avoid the infinite variance of the \textcolor{black}{L\'evy-stable} distribution function, several truncations are proposed. The truncation was justified by the observed change of slope of the tails on extensive datasets \citep{podobnik2001time}. For example, \textcolor{black}{when} evaluating the returns per minute of S\&P 500 index data \textcolor{black}{over the} ten \textcolor{black}{year} period \textcolor{black}{from 1985-95} a change of slope from $\alpha=1.4$ to $\alpha=3$ was found. The truncations make the variance finite, consequently the distribution function of the sum of independent random variables  converges to the normal distribution due to the central limit theorem for \textcolor{black}{large $N$}.  Nevertheless, a time series in some stock \textcolor{black}{market indices} can exhibit infinite variance, one \textcolor{black}{such case} is the variance of price fluctuations in Shanghai stock market index, which increases when the time frame is enlarged \citep{huang1995asian,lee2001stock}.\par
The aim of this paper is to formulate a uniform analytical approximation for the \textcolor{black}{L\'evy-stable} distribution function based on a series expansion. To achieve this aim we propose several regularizations of the inner and outer series expansions to ensure convergence. This will be an important tool to get the most accurate approximation reducing numerical errors (oscillations) when the \textcolor{black}{L\'evy-stable} function is evaluated.\par

This paper is divided in two parts. The first part introduces the \textcolor{black}{L\'evy-stable} distribution and the trans-stable function. They are defined by Fourier transformations in sections \ref{sec:Stable} and \ref{sec:trans} respectively.  The trans-stable function is shown to be identical to the \textcolor{black}{L\'evy-stable} distribution for $\alpha<1$ and it has the same asymptotic behaviour for  $\alpha>1$ for large events. The second part refers to section \ref{sec:Asymptotic} and it deals with the \textcolor{black}{closed} form--- analytical approximations--- of the \textcolor{black}{L\'evy-stable} distribution. For this purpose, two types of approximations are developed.
The first approximation refers to the inner expansion that converges asymptotically to the \textcolor{black}{L\'evy-stable} distribution as $x\rightarrow0$. The second approximation refers to the outer expansion that converges asymptotically as $x\rightarrow\infty$. For the outer expansion two cases are presented, one is obtained from the \textcolor{black}{L\'evy-stable} function in subsection \ref{OuterSolution} and the second one from \textcolor{black}{the} trans-stable function in subsection \ref{Outersolution2}.
Finally, the uniform solution in section \ref{sec:Uniform} is proposed as a result of matching the inner and the outer solution. The analytical equation of uniform solution proposed in this paper gives an approximated solution of the \textcolor{black}{L\'evy-stable} distribution function over the range $-\infty<x<\infty$.

\section{\label{sec:Central Limit theorem} Central Limit Theorem for L\'evy-stable Flights}
Section 35 of the book by Gnedenko and Kolmogorov \citep{GnedenkoKolmogorov1954} shows that the normal distribution is an ``attractor" of distributions with finite variances. On the other hand, the attractor of \textcolor{black}{power law distributions with infinite variances} corresponds to the more general ``\textcolor{black}{L\'evy-stable} law". \textcolor{black}{In other words the L\'evy-stable is a specific function \textcolor{black}{to which} other distributions converge}.\par
The fundamental concept of attractors is formulated as follows\textcolor{red}{.} If a normalized sum of a set of independent, identically distributed random variables $\left \{ X_{1}, X_{2}, X_{3}.... X_{N}\right \}$ satisfies:
\begin{equation}\label{eq:1}
 \lim_{N\rightarrow \infty }\frac{1}{\sigma_{N}}\left ({\sum_{i=1}^{N}X_{i}-\mu_{N} }  \right )=X,
\end{equation}
\textcolor{black}{then} $X$ belongs to the stable law. The coefficients $\mu_{N}$ and $\sigma_{N}$ represent the centering and normalizing values respectively \citep{GnedenkoKolmogorov1954}. \par
The Gnedenko-Kolmogorov theorem is a generalization of the classical central limit theorem which states that normalized sum of independent random variables with finite variance in Eq.~(\ref{eq:1}) converges to a variable that is normally distributed \citep{GnedenkoKolmogorov1954,Breiman1968}. This is the case of distributions with power-law tails \textcolor{black}{($\alpha \geq 2$)} with finite variance. The  normalized coefficient is $\sigma_{N}=\sqrt{N}$ and the centering coefficient is $\mu_{N}=NE[X]$, where $N$ represents the length of the sum and $E[X]$ refers to expected value \citep{mandelbrot1963variation,Uchaikin2011}. On the other hand, for independent random variables \textcolor{black}{power law distributions with infinite variances} \footnotemark[1]
\footnotetext[1]{Note: Infinite variance is observed for $0<\alpha<2$. This characteristic occurs for $0<\alpha\leq1$, as a consequence of not having a well-defined expected value $E\left [ X \right ]$. For $1<\alpha<2$, the integral in  the variance definition diverges  \citep{Uchaikin2011,zolotarev1986one,Feller1950}.} $0<\alpha<2$, Uchaikin and Zorotalev \citep{Uchaikin2011,zolotarev1986one} show that $X$ in Eq.~(\ref{eq:1}) follows a symmetric \textcolor{black}{L\'evy-stable} law if the normalization coefficient is $\sigma_{N}=N^{1/\alpha}$ and the centering coefficient is $\mu_{N}=0$ for $\alpha\leq1$  or $\mu_{N}=NE[X]$ for $\alpha>1$.\par 
\textcolor{black}{L\'evy-stable distributions belong to a wider class of infinitely divisible distributions (ID). A random variable $X$ is ID if it can be represented as the sum of a number $N$ of independent and identically distributed random variables \textcolor{black}{with a common law $(N)$} \citep{applebaum2009levy}, i.e.:
\begin{equation}\label{eq:1.0}
X\overset{}{=}\sum_{i=1}^{N}Y_{i}^{(N)} \quad \forall \quad N \in \mathbb{N}
\end{equation}}
\textcolor{black}{The pdf of $X$ is $f(x)$. If $f(x)$ is L\'evy-stable, then $f(x)$ is an ID distribution function.
 The proof of this statement is obtained by replacing $Y_{i}^{(N)}=(X_{i}-\mu_{N}))/\sigma_{N}$ into Eq.~(\ref{eq:1.0}). Then, by applying the limit  $N\rightarrow\infty$ Eq.~(\ref{eq:1}) is obtained. An equivalent definition of ID can be given in terms of the characteristic function. The characteristic function is defined as the Fourier transform of the probability density function $f(x)$,}
 \textcolor{black}{
 \begin{equation}\label{eq:1.1}
 \varphi(t)=\mathbb{E}(e^{itx})= \int_{-\infty }^{\infty} f(x)e^{itx}dx.
 \end{equation}}

 \textcolor{black}{
The characteristic function of the ID distribution can be derived as follows. Consider $X$ as a sum of two independent random variables $X=Y_{1}+Y_{2}$ with pdf's $f_{1}(x)$ and $f_{2}(x)$ respectively.}
\textcolor{black}{For the convolution of the two probability distributions \citep{vasconcelos2004guided}, the pdf of X has the form,}
\textcolor{black}{
\begin{equation}\label{eq:1.2}
f(x)=\int_{-\infty }^{\infty }f_{1}(k)f_{2}(x-k)dk.
\end{equation}}
\textcolor{black}{By substituting Eq.~(\ref{eq:1.2}) into Eq.~(\ref{eq:1.1}) and interchanging the order of the integration the equation for the characteristic function of $X$ is obtained,}
\textcolor{black}{
\begin{equation}\label{eq:1.3}
 \varphi_{X}(t)=\varphi_{Y_{1}}(t)\varphi_{Y_{2}}(t).
\end{equation}}
\textcolor{black}{Assuming that $Y_{1}$ and $Y_{2}$ are identically distributed, the characteristic function of  $f(x)$ can be defined as $ \varphi_{X}(t,2)=(\varphi(t))^{2}$. In general, for the sum of $N$  independent and identically distributed  random variables in Eq.~(\ref{eq:1.0}), the characteristic function is given by:}

\textcolor{black}{
\begin{equation}\label{eq:1.4} 
\varphi_{X}(t,N)=(\varphi_{N}(t))^{N}.
\end{equation}}
\textcolor{black}{Consequently, Eq.~(\ref{eq:1.0}) and  Eq.~(\ref{eq:1.4}) are equivalent. Then, the limit is applied in  Eq.~(\ref{eq:1.4}), $\varphi_{X}(t)=\lim_{N\rightarrow\infty  }\varphi_{X}(t,N) $. \textcolor{black}{As a consequence, $\varphi_{X}(t)$ is the characteristic function of the pdf of the random variable $X$. 
This statement constitutes the Levy continuity theorem that guarantees pointwise convergence \citep{applebaum2009levy,Bertoin1996}}. 
The Lévy-Khintchine formula or Triple Lévy gives the general equation for ID distributions \citep{applebaum2009levy}. This formula determines the class of characteristic function where the pdf is calculated from its Fourier transform \citep{samuels1975infinitely,klenke2014infinitely,Bertoin1996,Samorodnitsky1994}.
The L\'evy-stable distribution constitutes a special case of the general Lévy-Khintchine in one-dimensional case that is presented in the next section \citep{Bertoin1996}.}\par

\par

\section{\label{sec:Stable} L\'evy-stable Distribution Function}
\textcolor{black}{The \textcolor{black}{L\'evy-stable} distribution is given by the Fourier transform of Eq.~(\ref{eq:1.1}),}
\begin{equation}\label{eq:SDor}
{{f}}(x;\alpha ,\beta ,\sigma,\mu)=\frac{1}{2\pi }\int\limits_{-\infty }^{\infty }{\varphi (t;\alpha ,\beta ,\sigma,\mu){{e}^{ixt}}dt}.
\end{equation}
\textcolor{black}{Where $\varphi (t)$ is presented in Section 34 of the Gnedenko-Kolmogorov book \citep{GnedenkoKolmogorov1954} as,}

\begin{equation}\label{eq:SDcha}
\varphi (t;\alpha ,\beta ,\sigma,\mu )={{e}^{(it\mu -{{\left| {\sigma}t \right|}^{\alpha }}(1-i\beta sgn(t)\Phi ))}}.
\end{equation}
 
The four parameters involved are: the stability parameter \textcolor{black}{$\alpha \in ( 0 \left. 2 \right]$}, the skewness parameter $\beta \in \left[ -1\left. +1 \right] \right.$, the scale parameter ${\sigma}\in \left( 0\left. +\infty  \right) \right.$, and the location parameter $\mu \in \left( -\infty +\infty  \right)$. The parameter $\alpha$ constitutes the characteristic exponent of the asymptotic power-law in the tails and it determines whether the mean value and the variance exist. The \textcolor{black}{L\'evy-stable} distribution with ${0< \alpha \leq  1}$ does not have a mean value and it has a define variance only for $\alpha=2$ \citep{arnold1978227}.
\begin{flushleft}
The function $sgn\left ( t \right )$ represents the sign of $t$ and the function $\Phi $ is defined as:
\begin{equation}\label{eq:SDfi}
\Phi =\left\{ \begin{matrix}
   \tan \left( \frac{\pi \alpha }{2} \right) & \alpha \ne 1,  \\
   -\frac{2}{\pi }\log \left| t \right| & \alpha =1.  \\
\end{matrix} \right.
\end{equation}
\end{flushleft}
The \textcolor{black}{L\'evy-stable} distribution is the family of all attractors of normalized sums of independent and identically distributed random variables. The most well-known \textcolor{black}{L\'evy-stable} distribution functions are the Cauchy distribution with $\alpha=1$ and the normal distribution function  with $\alpha=2$. Both functions have $\beta=0$, which means they are  symmetric distributions about their mean \citep{zolotarev1986one}.\par
In this paper we will focus on \textcolor{black}{symmetric distributions}. For this case the \textcolor{black}{L\'evy-stable} distribution can be normalized as follows:
\begin{equation}
{{f}}(x;\alpha,\beta=0,\sigma,\mu)=Re\left \{ S\left ( \frac{x-\mu }{\sigma} ,\alpha \right ) \right \},
\end{equation}
where the general distribution function is given by the following equation:
\begin{equation}\label{eq:SD}
S( x;\alpha)=\frac{1}{\pi }\int\limits_{0}^{\infty }{{{e}^{-t}}^{^{\alpha }}{{e}^{ixt}}dt}.
\end{equation}
The real part of this function corresponds to the normalized \textcolor{black}{L\'evy-stable} distribution,
\begin{equation}\nonumber
s(x;\alpha)=Re(S(x;\alpha)).
\end{equation}
Consequently, \textcolor{black}{by applying Euler's formula we arrive at} \citep{Weisstein1999}:

\begin{equation}\label{eq:SD_1}
s(x;\alpha)=\frac{1}{\pi }\int\limits_{0}^{\infty }{{{e}^{-t}}^{^{\alpha }}\cos (tx)dt}.
\end{equation}

\section{\label{sec:trans}Trans-Stable Function}
Zolotarev (1986) used the term ``trans-stable" to refer to a power series expansion that converges to the \textcolor{black}{L\'evy-stable} distribution for $0<\alpha<1$ only \citep{zolotarev1986one}.
In this paper, trans-stable is the function which one of its solutions originates Zolotarev series when the series expansions are applied around $x\rightarrow \infty $. First we define  \textit{the complex trans-stable function} in the range of $0<\alpha<2$. For $\alpha<1$, the \textcolor{black}{L\'evy-stable} distribution and the trans-stable function are identical. For $\alpha>1$, the trans-stable function and the \textcolor{black}{L\'evy-stable} distribution  present the same asymptotic behaviour for $x\rightarrow \infty$.
Consequently, our trans-stable function can be used to find a numerical approximation of the \textcolor{black}{L\'evy-stable} distribution function for $\alpha>1$ for large events.\par
First,  \textit{the complex trans-stable function}  is defined as an integral over the path $C$ in the complex plane:
\begin{equation}\label{eq:C_TS}
G_{C}\left (x;\alpha \right )=\frac{1}{\pi }\int_{C}I\left ( x,z;\alpha\right ) dz,
 \end{equation}
 where
\begin{equation}\label{eq:I_TS}
I\left ( x,z;\alpha\right )= {e^{-z}}^{\alpha} e^{ixz}.
\end{equation}
The relation of this function to the \textcolor{black}{L\'evy-stable} $S(x;\alpha)$ and the trans-stable $T(x;\alpha)$ functions is obtained by choosing a particular path $C$ in the complex plane. Then, the \textcolor{black}{L\'evy-stable} distribution and  trans-stable function are given by Eq.~(\ref{eq:SD2})  and (\ref{eq:TS}) respectively:
\begin{equation}\label{eq:SD2}
S(x;\alpha)=G_{\left [0,\infty \right )}(x;\alpha)=\frac{1}{\pi }\int\limits_{0}^{\infty }{{{e}^{-t}}^{^{\alpha }}{{e}^{ixt}}dt},
\end{equation}
\begin{equation}\label{eq:TS}
T(x;\alpha)=G_{\left [0,i\infty \right )}(x;\alpha)=\frac{1}{\pi } \int\limits_{0}^{i\infty }{{{e}^{-{{t}^{\alpha }}}}{{e}^{ixt}}dt}.
\end{equation}

First it will be shown that for $0< \alpha \leq1$, both \textcolor{black}{L\'evy-stable} $S(x;\alpha)$ and trans-stable $T(x;\alpha)$ functions are identical. For $1<\alpha<2$ it will be demonstrated that both functions exhibit the same asymptotic behaviour when $x\rightarrow \infty $.\\

This demonstration is based on the evaluation of the \textit{complex trans-stable integral} Eq.~(\ref{eq:C_TS}) using polar representation for $\alpha\leq1$ and rectangular representation for $\alpha>1$ on the complex integrand. The demonstrations are presented in the following subsections. 

\begin{figure} [htbp]
\centering
\includegraphics[angle=0,scale=0.33,trim=0cm 3cm 4cm 1.8cm]{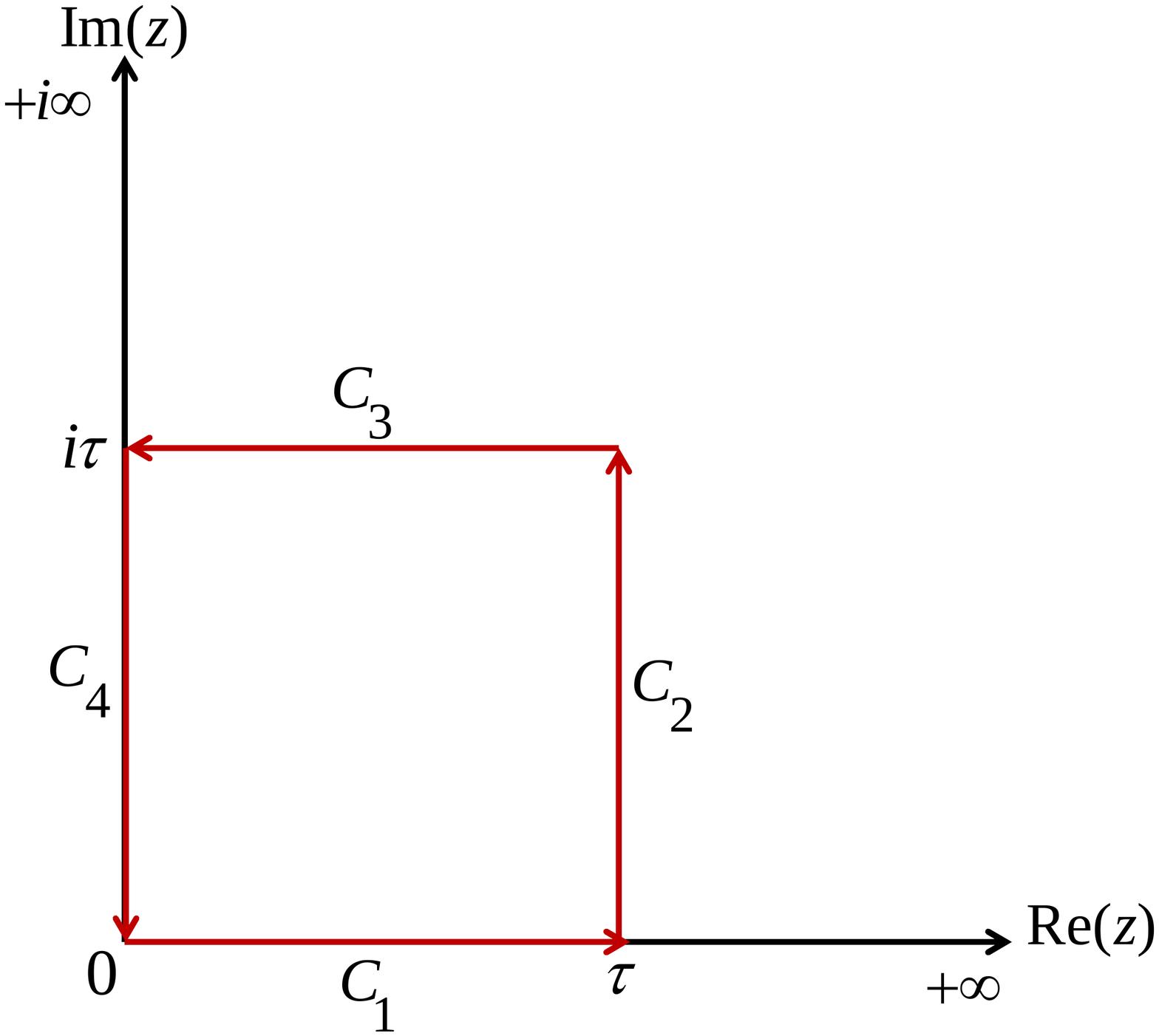}
\caption{ \textit{Contour integration for} Eq.~(\ref{eq:C_TS})} 
\label{fig:contour}
\end{figure}

\subsubsection{\label{subsec:transA} For ${0< \alpha \leq  1}$}

Here we will show that for $0< \alpha \leq1$ the \textcolor{black}{L\'evy-stable} and trans-stable functions are identical. This demonstration will be done by considering the closed contour shown in Figure~\ref{fig:contour}. Since the complex function in Eq.~(\ref{eq:I_TS}) is analytical over the complex plane, the integral over the closed contour Eq.~(\ref{eq:C_TS}) is zero,

\begin{equation}
\oint I\left ( x,z;\alpha\right ) dz=0.
\end{equation}

Let us take the contour in Figure~\ref{fig:contour} that can be divided into four straight paths so that:

\begin{equation}\label{eq:SUM_C_TS}
\sum_{1}^{4}\int_{C_{i}}I\left ( x,z;\alpha\right ) dz=0.
\end{equation}

Now, we will take the limit when $\tau\rightarrow \infty $ in Figure~\ref{fig:contour}. Using Eq.~(\ref{eq:SD2},\ref{eq:TS},\ref{eq:SUM_C_TS}) the following equation is obtained:

\begin{equation}\label{eq:SUM2_C_TS}
S( x;\alpha)-T(x;\alpha)=-{\lim }_{\tau\to \infty }\sum_{i=2}^{3}\int_{C{i}}I\left ( x,z;\alpha\right ) dz.
\end{equation}

To evaluate the right hand side in Eq.~(\ref{eq:SUM2_C_TS}) it is convenient to use the polar representation of the complex number ${z=re^{i\theta}}$ and express Eq.~(\ref{eq:I_TS}) in polar coordinates:

\begin{equation} \label{eq:P_I_TS}
I\left ( x,z;\alpha\right)= {e^{g(x,r,\theta;\alpha)+ih(x,r,\theta;\alpha)}},
\end{equation}
\begin{equation}\label{eq:g1}
g(x,r,\theta;\alpha)={-{{r}^{\alpha }}\cos \left( \theta \alpha  \right)-rx\sin \theta },
\end{equation}
\begin{equation}\label{eq:h1}
h(x,r,\theta;\alpha)= {-{{r}^{\alpha }}\sin \left( \theta \alpha  \right)+rx\cos \theta  }.
\end{equation}
It will be adopted the nomenclature of signal theory, where the polar notation separates the effects of instantaneous amplitude $\arrowvert I\arrowvert=e^{g}$ 
and its instantaneous phase 
$h$ of a complex function \cite{feldman2011hilbert}.
Consequently, $g(x,r,\theta;\alpha)$ represents the \textit{attenuation factor} and $h(x,r,\theta;\alpha)$ represents the \textit{oscillation factor}.

Now let us notice that ${\lim }_{r\to \infty } g(x,r,\theta;\alpha)=-\infty $ for $0< \alpha \leq1$ at any value of $x$. This statement is based on the fact that $cos(\theta\alpha)\geq0$ in the first quadrant for $\alpha\leq1$. Consequently, ${\lim }_{r\to \infty } I\left ( x,z;\alpha\right )=0$ so that the integral of the right side of the Eq.~(\ref{eq:SUM2_C_TS}) vanishes at $\tau\to \infty$, therefore:
\begin{equation}\label{eq:S=T}
S( x;\alpha)=T(x;\alpha)\quad \textrm{if} \quad 0< \alpha \leq1.
\end{equation}
So, the Eq.~(\ref{eq:S=T}) will allow to use the trans-stable function $T(x;\alpha)$ instead of the \textcolor{black}{L\'evy-stable} distribution function $S(x;\alpha)$ for $0< \alpha \leq  1$ in the numerical integration. This is with the aim to remove numerical oscillation, specifically in the tails.
It is noticeable that the integration of the trans-stable function $T(x;\alpha)$ in Eq.~(\ref{eq:TS})is performed over the imaginary axis. Applying the following change of variable $t\rightarrow-it$ (formally done by defining $u=-it$ so that $du = -idt$ and later replacing the dummy variable $u$ by $t$ inside the integral), the trans-stable function is converted into a Laplace transformation. Consequently, the integration is performed over the real axis. The Fourier and Laplace representations for $T(x;\alpha)$ are shown in Eq.~(\ref{eq:Trans-fourier and Laplace}),
\begin{equation}\label{eq:Trans-fourier and Laplace}
T(x;\alpha)=\frac{1}{\pi }\int\limits_{0}^{i\infty }{{{e}^{-{{t}^{\alpha }}}}{{e}^{ixt}}dt}=\frac{1}{\pi }\int\limits_{0}^{\infty }{{{e}^{-{{(it)}^{\alpha }}}}{{e}^{-xt}}idt}.
\end{equation}

 Figure~\ref{Fig01C_StableDiscontinuties}  compares the Fourier representation of the \textcolor{black}{L\'evy-stable} distribution function $S(x;\alpha)$ and the Laplace representation of trans-stable function $T(x;\alpha)$. The integration is performed using a recursive adaptive Simpson quadrature method \cite{gander2000adaptive}. It is evident that the  Laplace representation removes the oscillations of the Fourier representation of the \textcolor{black}{L\'evy-stable} distribution for $\alpha<1$.  

It is important to add that the \textcolor{black}{L\'evy-stable} distribution function and trans-stable function hold the same value for their Fourier and Laplace transform representations. The difference between each transform representation is the axis in which each function is integrated. The expressions are shown in Table \ref{tab:1a}.\pagebreak
\begin{figure} [H]\nonumber
\centering
\includegraphics[scale=0.21,trim=2cm 2cm 0cm 1.5cm]{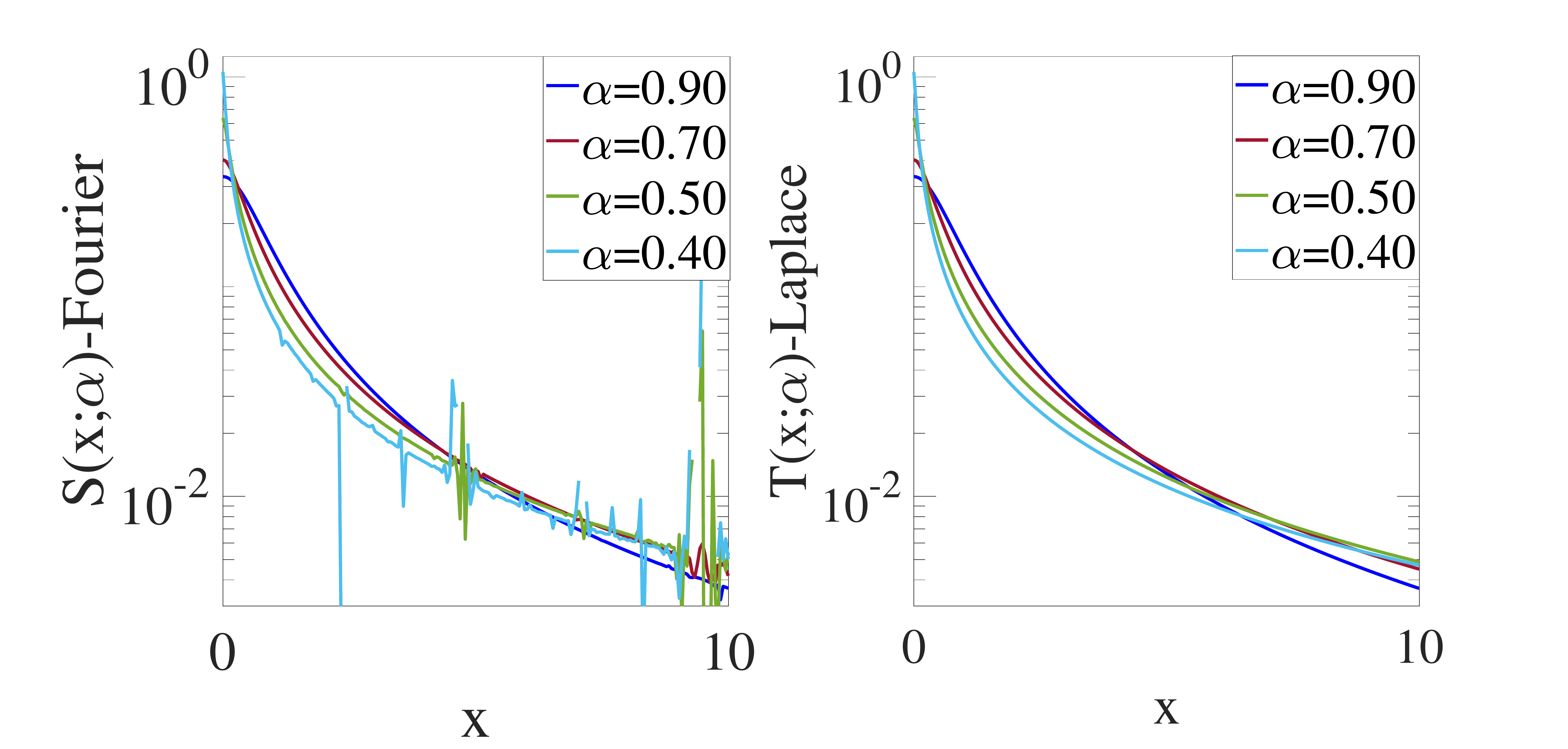}
\end{figure}
\begin{figure} [H]
\centering
\includegraphics[scale=0.21,trim=2cm 1cm 0cm 2.0cm]{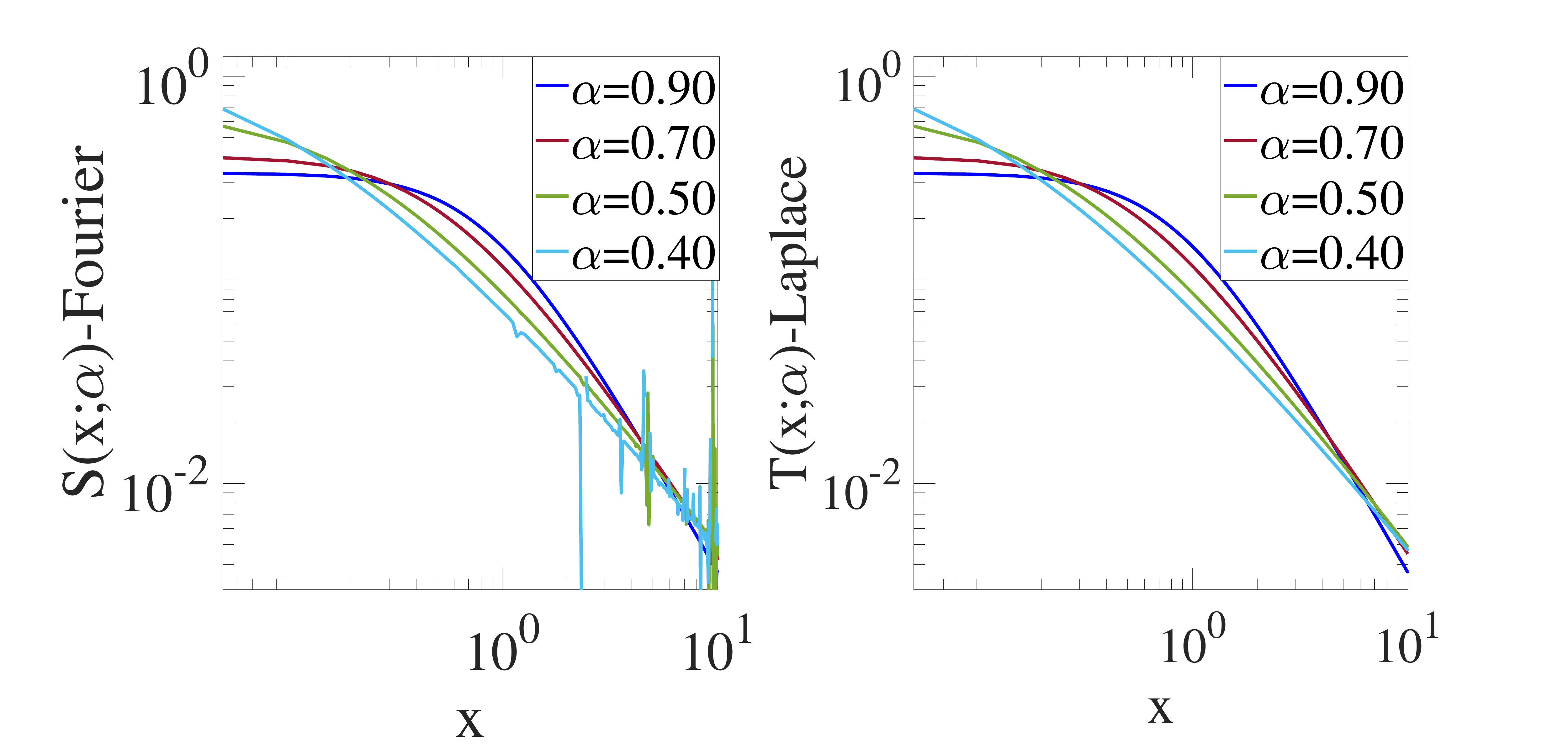}
\caption{\textit{Comparison of numerical integration $0<\alpha<1$ between Fourier and Laplace transform of the \textcolor{black}{L\'evy-stable} $S(x;\alpha)$ and the trans-stable $T(x;\alpha)$ functions using recursive adaptive Simpson quadrature method \cite{gander2000adaptive}. The absolute error tolerance of the method is $\xi=3.5 \times 10^{-8}$. Top plots are shown in semi-logarithmic scale and in logarithmic scale.}}
\label{Fig01C_StableDiscontinuties}
\end{figure}
\begin{table}[htbp!]
\begin{tabular}{  | M{5em} | M{3.2cm} | M{3.5cm} |} 
\hline
{}& \textcolor{black}{L\'evy-stable} distribution  $S(x;\alpha)$ & Trans-stable function $T(x;\alpha)$ \\ 
\hline
Fourier transform &{$ \frac{1}{\pi }\int\limits_{0}^{\infty }{{{e}^{-{{t}^{\alpha }}}}{{e}^{ixt}}dt}
  $  } & ${\frac{1}{\pi }\int\limits_{0}^{i\infty }{{{e}^{-{{t}^{\alpha }}}}{{e}^{ixt}}dt}}$ \\ 
\hline
Laplace transform & $\frac{1}{\pi }\int\limits_{0}^{-i\infty }{{{e}^{-{{\left( it \right)}^{\alpha }}}}{{e}^{-xt}}idt}$ & $\frac{1}{\pi }\int\limits_{0}^{\infty }{{{e}^{-{{(it)}^{\alpha }}}}{{e}^{-xt}}idt}$ \\ 
\hline
\end{tabular}
 \caption {\textit{Summary of Fourier and Laplace representations for the \textcolor{black}{L\'evy-stable}  and the trans-stable functions}}
   \label{tab:1a}
\end{table}

\subsubsection{\label{subsec:transB} For ${1<\alpha <2}$ }

Here we will shown that for $1< \alpha <2$ the \textcolor{black}{L\'evy-stable} and trans-stable functions have the same asymptotic behaviour on large events if the integrals are appropriately truncated.\par 
Let us recall Eq.~(\ref{eq:g1}) for the \textit{attenuation factor},
\begin{equation}\nonumber
g(x,r,\theta;\alpha)={-{{r}^{\alpha }}\cos \left( \theta \alpha  \right)-rx\sin \theta }.
\end{equation}
In the previous section, it was shown that $cos(\theta\alpha)$ is always positive in the first quadrant of the complex plane if $0<\alpha\leq1$. Otherwise, if $\alpha>1$, then $cos(\theta\alpha)<0$ when $ \theta=\pi/2$. Consequently, ${\lim }_{r\to \infty } I(x,r,\theta;\alpha)=\infty $ in this range, so that the right hand side of Eq.~(\ref{eq:SUM2_C_TS}) can not be neglected. Therefore $S(x) \neq T(x)$ if $\alpha >1$.\par
We can find an approximation between these two functions if the $\tau$ value in the contour of Figure~\ref{fig:contour} is kept large but finite $(\tau<\infty)$. Thus, Eq.~(\ref{eq:SUM_C_TS}) becomes:

\begin{equation}\label{eq:SUM3_C_TS}
S( x;\alpha,\tau)-T(x;\alpha,\tau)=-\sum_{i=2}^{3}\int_{S{i}}I\left ( x,z;\alpha\right ) dz,
\end{equation}
where $S(x; \alpha,\tau)$ and $T(x; \alpha,\tau)$ are the truncated integrals in Eqs.~(\ref{eq:SD2}) and (\ref{eq:TS}) respectively:

\begin{equation}\label{eq:S3}
S(x;\alpha,\tau)= \frac{1}{\pi }\int\limits_{0}^{\tau}{{{e}^{-t}}^{^{\alpha }}{{e}^{ixt}}dt},
\end{equation}

\begin{equation}
T(x;\alpha,\tau)= \frac{1}{\pi } \int\limits_{0}^{i\tau}{{{e}^{-{{t}^{\alpha }}}}{{e}^{ixt}}dt}.
\end{equation}

Now, the right hand of Eq.~(\ref{eq:SUM3_C_TS}) can be evaluated in the limit where $x\rightarrow\infty$. First, notice that in the contour of integration in Figure~\ref{fig:contour} the magnitude of $r$ is bounded by the condition $0<r<\sqrt{2}~\tau$ and $sin(\theta)>0$ in the first quadrant, thus:
\begin{equation}\nonumber
{\lim }_{x\to \infty } g(x,r,\theta;\alpha)=-\infty. 
\end{equation}
Consequently, ${\lim }_{x\to \infty }I\left ( x,z;\alpha\right )=0$ so that the integral on the right of Eq.~(\ref{eq:SUM3_C_TS}) vanishes at $x\rightarrow\infty$. Therefore, the asymptotic behaviour is obtained for $1< \alpha <2$, 
\begin{equation}\label{eq:S=TS}
S( x;\alpha,\tau)\sim T(x;\alpha,\tau)\quad \textrm{as} \quad x \rightarrow \infty.
\end{equation}
This demonstrates that both functions are  asymptotically equivalent when the integrals are truncated.\par
\begin{figure} [H]
 \centering
 \includegraphics[angle=0,scale=0.42,trim=2cm 2cm 2cm 2cm]{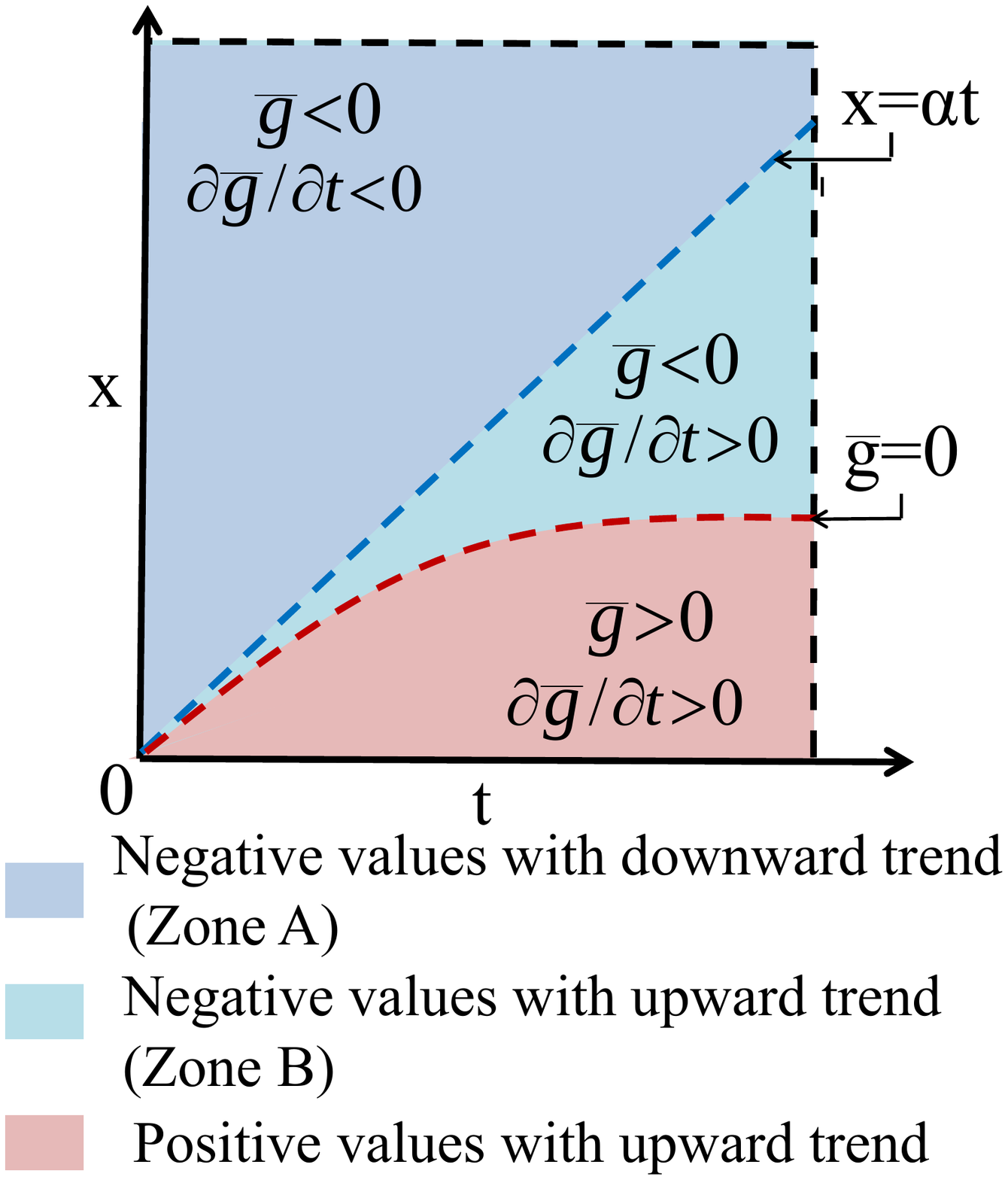}
 \caption{ \textcolor{black}{\textit{Curve $\bar{g}=0$ separates two regions with $\bar{g}>0$ and $\bar{g}<0$. In the latter region, two zones can be distinguished: zone A with $ \partial g/ \partial t <0$ and zone B with $ \partial g/ \partial t >0$. The equation $x=\alpha t$ is an estimation of the boundary  between zones A and B.}}}  
 \label{fig:Fig01B_GandI}
 \end{figure}

The next step is to find the truncation value $\tau$ that leads to the best approximation of these functions. The value of $\tau$ should be chosen to minimize the truncation error and at the same time to make the domain of integration as small as possible. With this aim, the trans-stable function $T(x;\alpha,\tau)$ in Eq.~(\ref{eq:TS}) is expressed in its Laplace representation by using the change of variable $t\rightarrow-it$. Thus,

\begin{equation}\label{eq:TS3}
T(x;\alpha,\tau) =\frac{1}{\pi }\int_{0}^{\tau}\bar{I}\left(x,t;\alpha\right )dt,
 \end{equation}
 where $\bar{I}$ corresponds to Laplace transform integrand shown in Eq.~(\ref{eq:Trans-fourier and Laplace}) and Table \ref{tab:1a},
 \begin{equation}\label{eq:I_TS2}
 \bar{I}\left ( x,t;\alpha\right )= {e^{\left (-it \right )}}^{\alpha} e^{-xt}i.
 \end{equation}
Then, considering  Euler's representation for a complex exponential function $e^{i\theta}=cos\left (\theta\right )+isin\left (\theta\right )$, the following equations are obtained to express Eq.~(\ref{eq:I_TS2}):
 \begin{equation} \label{eq:P_I_TS2}
 \bar{I}\left ( x,t;\alpha\right)= {e^{\bar{g}(x,t;\alpha)+i\bar{h}(x,t;\alpha)}},
 \end{equation}
 
 \begin{equation}\label{eq:P_I_TS2A}
 \bar{g}(x,t;\alpha)=-{{t}^{\alpha }}\cos \left( \frac{\pi \alpha }{2} \right)-xt,
 \end{equation}
 \begin{equation}\label{eq:P_I_TS2B}
 \bar{h}(x,t;\alpha)= {-{{t}^{\alpha }}\sin \left( \frac{\pi \alpha}{2}   \right)+ \frac{\pi}{2}  }.
 \end{equation}
The instantaneous amplitude $|\bar{I}|=e^{\bar{g}}$ will be determined by the \textit{attenuation factor} in Eq.~(\ref{eq:P_I_TS2A}). For that reason, an analysis of $\bar{g}(x,t;\alpha)$ was made in Figure \ref{fig:Fig01B_GandI}. The curve $\bar{g}(x,t;\alpha)=0$ divides two regions, one with exponential growth $(\bar{g}>0)$ and the other with exponential decay $(\bar{g}<0)$.\par
\begin{figure*} [htbp]
\centering
\includegraphics[scale=0.44,trim=5.5cm 1.0cm 1cm 2cm]{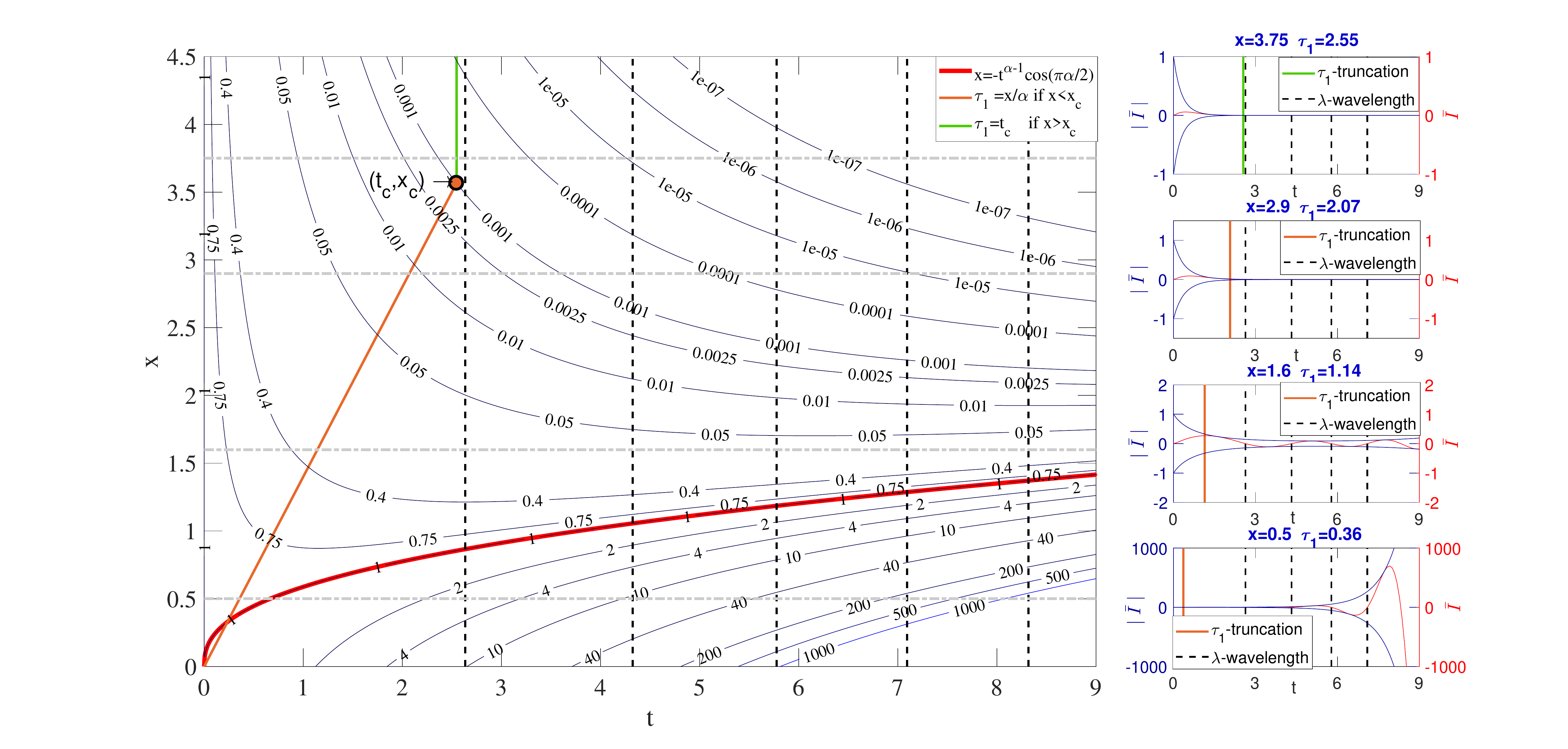}
\caption{ \textit{Contour plot of } $e^{\bar{g}}$ \textit{that represents the instantaneous amplitude $|\bar{I}|$ for $\alpha=1.4$} \textit{and tolerance $\epsilon  =10^{-3}$ in Eq.~(\ref{eq:P_I_TS2}) and (\ref{eq:P_I_TS2A})}. \textit{The red line $x$ was drawn by considering $\bar{g}=0$ which represents the limit between the positive and negative values of the attenuation factor} $\bar{g}$.  \textit{ The truncation $\tau_{1} $ is applied following Eq.~(\ref{eq:TS2a}). Downward trend before $\tau_{1}$ and upward trend after $\tau_{1}$ of  ${\rvert\bar{I}\rvert}$  can be observed in the right part of the figure for $x=0.5$ and $x=1.6$. For $x=2.9$ and $x=3.8$ the truncation is made after reaching a small value on the modulation ${\bar{I}}$.}}
\label{fig:Fig02_contourtranstable}
\end{figure*}
In Figure \ref{fig:Fig01B_GandI}, two sub-regions can be recognized in $\bar{g}<0$. The first one, ``Zone A'' which contains negative $\bar{g}$ values with downward trend $ \partial g/ \partial t <0$ that is faster as  $x\rightarrow\infty$. The second sub-region is ``Zone B'', it contains smaller negative  $\bar{g}$ values  that  follow an upward trend and $ \partial g/ \partial t >0$ displaying an increase behaviour when $x\rightarrow0$. Considering these sub-regions, the truncation $\tau$ in Eq.~(\ref{eq:TS3}) will depend on $x$ value as follows:
\begin{itemize}
\item For $x\rightarrow0$, The integration must avoid zone B. The values of $\bar{g}(x,t;\alpha)$ in this zone lead to an exponential growth due to an upward trend $\partial g/ \partial t >0$, consequently $|\bar{I}|\nrightarrow0$. 
\item For $x\rightarrow\infty$, the integration should be restricted to zone A. The downward trend $ \partial g/ \partial t <0$ leads to obtain  $\bar{g}(x,t;\alpha)\rightarrow0$. Consequently, the convergence of $|\bar{I}|\rightarrow0$ occurs faster as $t\rightarrow\infty$.
\end{itemize}

For $x\rightarrow0$, the cut off $\tau_{1}$  which avoids most of zone B is defined by $x=\alpha t$. This equation is an estimation of the boundary between zones A and B for all range of $\alpha$ values. 

The cut off $\tau_{1}$ obeys a linear equation and is obtained from the following equations: 

\begin{equation}\label{eq:bar{I}}
\begin{matrix}
{e^{\bar{g}(x,\tau_{1};\alpha)}=|\bar{I}|=\epsilon} & {\textrm{and} }&{{x}}=\alpha {{\tau_{1}}}, 
\end{matrix}
\end{equation}
where the tolerance $\epsilon$ represents a negligible instantaneous amplitude $\left| \bar{I}\right| $.\par

For $x\rightarrow\infty$, the cut off $\tau_{1}$ will restrict the integration of $\bar{I}$ on a closed interval $[0,t_{c}]$. This occurs due to a faster downward trend $ \partial g/ \partial t <0$. The $t_{c}$ value represents the point where the instantaneous amplitude can be considered a negligible quantity $|\bar{I}|=\epsilon$. Thus, the cut off $\tau_{1}$ obeys an equation of a vertical line  $\tau_{1}=t_{c}$.\par
Notice that there are two different definitions for $\tau_{1}$. Each one corresponds to a particular sub-regions A $(x\rightarrow\infty)$ or B $(x\rightarrow0)$. Consequently, the truncation $\tau_{1}$ for the trans-stable function is defined by two equations which depend on the $x$ and $\epsilon$ values. These two equation have their intersection point at $(t_c,x_{c})$:

\begin{equation} \label{eq:TS2a}
\tau_{1}(\epsilon,x) =\left\{\begin{matrix}
t_{c}(\epsilon)\quad if\quad x>x_{c}\\ 
{x}/{\alpha}\quad if\quad x<x_{c}
\end{matrix}\right.\quad for \quad\alpha>1,
\end{equation}
where $t_{c}(\epsilon)$ and $x_{c}$ are given by the implicit form of the following equations:
\begin{equation}
\begin{matrix}\label{eq:TS2b}
{\quad\alpha {{t}_{c}}^{2}+{{t}_{c}}^{\alpha }\cos \left( {\pi \alpha }/{2} \right)+\ln (\epsilon)=0},\\ {{{x}_{c}}=\alpha {{t}_{c}}}.
\end{matrix}
\end{equation}

Figure \ref{fig:Fig02_contourtranstable} illustrates the contour plot of the instantaneous amplitude $|\bar{I}|$ for $\alpha=1.4$. The truncation $\tau_{1}$ is presented as a cut-off made when a negligible value of instantaneous amplitude is achieved $|\bar{I}|=\epsilon=10^{-3}$. The point $(x_{c},t_{c})$ is located  at the intersection between the contour line of the given tolerance $\epsilon$ and the equation ${{\tau_{1}}}={{{x}}}/{\alpha} $. The truncation $\tau_{1} $ avoids zone B which contains negative values for $\bar{g}$ with $ \partial g/ \partial t >0$. One can observe that there is an abrupt upward trend in $ \rvert\bar{I}\rvert$ for $x\to 0$. So, the truncation $\tau_{1} $ allows us to make a perfect cut off before this upward trend starts. It is noticeable that with a small tolerance $\epsilon$ the intersection will occur in the rightmost part of the figure, consequently the interval of integration will be wider and a more accurate result can be obtained.\par 
Figure \ref{Fig02B_StablevsTranstable} shows how the solutions of trans-stable and \textcolor{black}{L\'evy-stable} distribution functions are quite similar after $x_c$ value, which depends on the tolerance $\epsilon$. For a smaller $\epsilon$ the similarity of both asymptotic series is expected to improve due to a wider interval of integration. However, the value $x_{c}$ will be higher and the similarity will start at the rightmost part of the axis.
 \begin{figure} [H]\nonumber
  \centering
  \includegraphics[scale=0.21,trim=4.5cm 2cm -2cm 4cm]{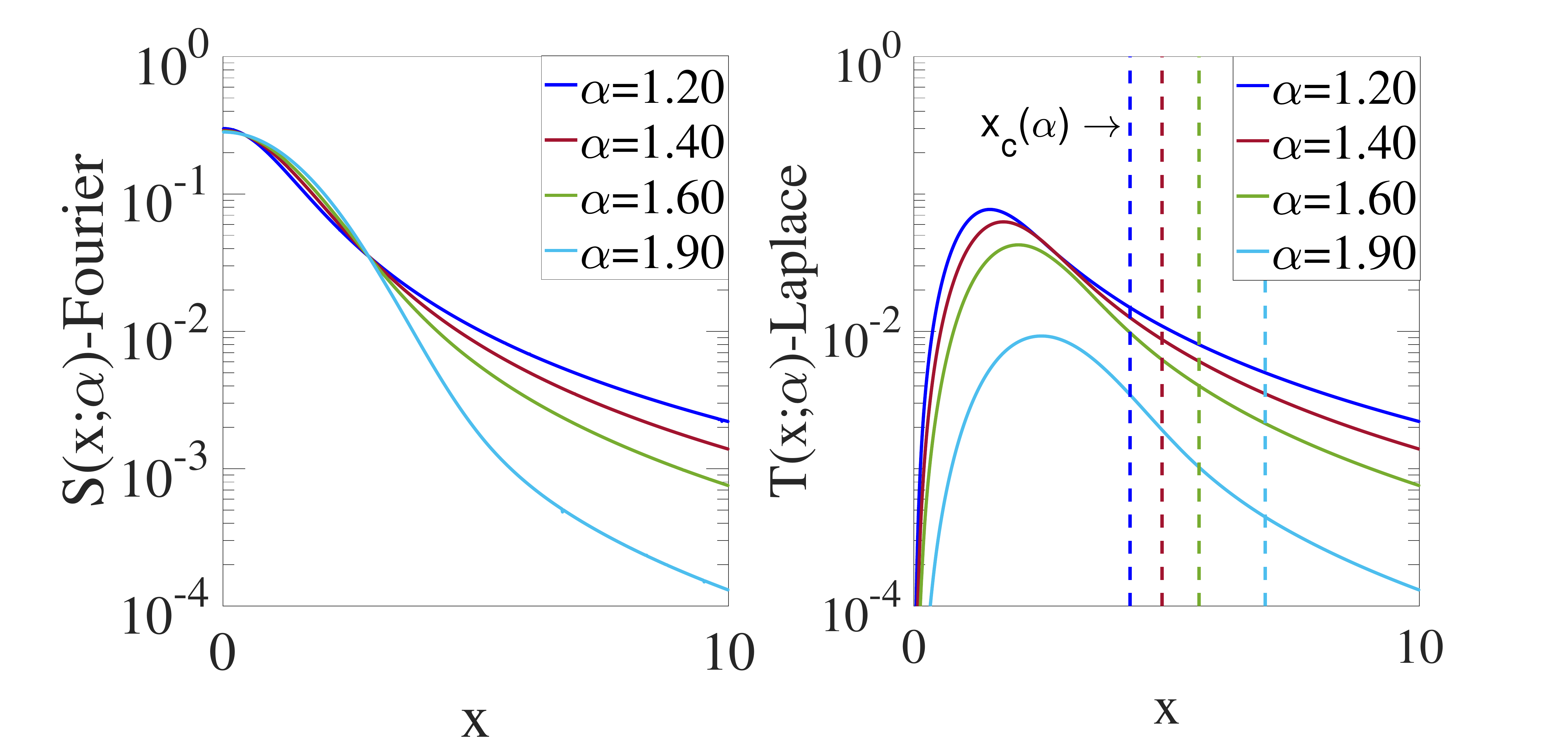}
  \end{figure}
  \begin{figure} [H]
  \centering
  \includegraphics[scale=0.21,trim=4.5cm 2cm 0cm 2cm]{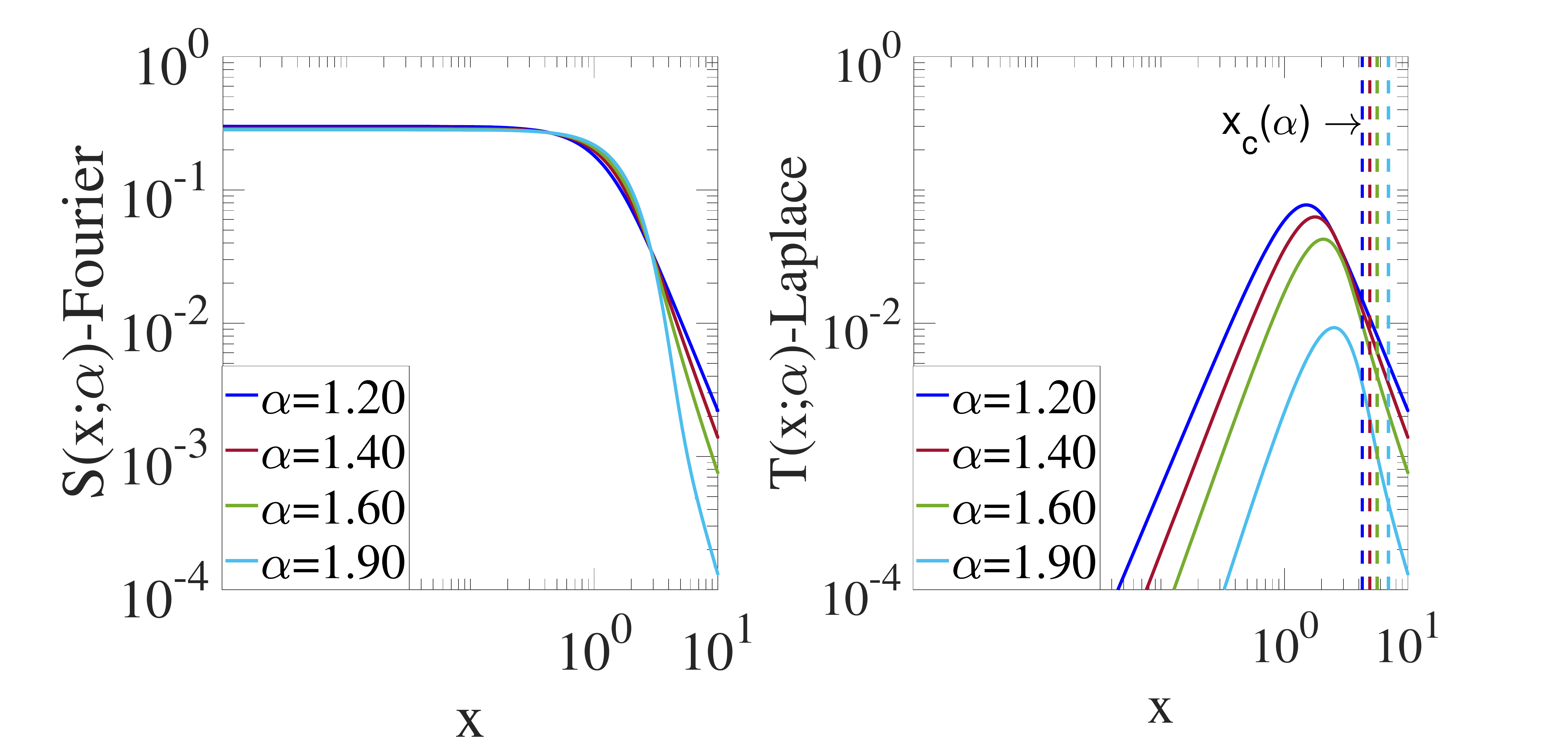}
  \caption{\textit{Comparison of numerical integration $1<\alpha<2$ between Fourier and Laplace transform of \textcolor{black}{L\'evy-stable} $S(x;\alpha)$ and trans-stable $T(x;\alpha)$ functions using recursive adaptive Simpson quadrature method \cite{gander2000adaptive}. The absolute error tolerance of the method is $\xi=3.5\times10^{-8}$. Top plots are shown in semi-logarithmic scale and in logarithmic scale. } }
  \label{Fig02B_StablevsTranstable}
  \end{figure}

\section{\label{sec:Asymptotic}Asymptotic Expansions}

Asymptotic expansions are developed to obtain closed-form representations for the \textcolor{black}{L\'evy-stable} distribution function $S(x,\alpha)$. These expansions are based on the Taylor series of the complex exponential function,\par
\begin{equation}\label{eq:Texp}
 {{e}^{z}}\sim\sum\limits_{k=0}^{n}{\frac{{{ z}^{k}}}{k!}}\quad as \quad n\rightarrow \infty.
\end{equation}
 Two different asymptotic expansions will be performed. The first one corresponds to the `inner expansion'. To get this solution the \textcolor{black}{L\'evy-stable} distribution function is evaluated by expanding  $e^{ixt}$ of Eq.~(\ref{eq:SD2}) and (\ref{eq:S3}) around $x=0$.
 The second one refers to the `outer expansion', which is the asymptotic series expansion for $x\rightarrow\infty$. 
 When $x>>1$, the oscillations of the integrands in Eq.~(\ref{eq:SD2}) and (\ref{eq:S3}) are large. Consequently, there are important cancellations due to factor $e^{ixt}$ in the integral.  Thus, we focus our integration in the region with the major contribution in the integral, that is around $t=0$. In consequence, the amplitude of the integral ${e^{t}}^{\alpha}$ is replaced by its Taylor expansion around $t=0$. 
 To guarantee the convergence of the series expansion, the improper integrals are truncated. The truncation occurs because of the sufficient conditions for Riemann integral existence. These conditions are that the integrand must be bounded and the domain of integration is a closed interval \cite{yoneda1981riemann,bartle1996return}.

\subsection{\label{InnerSolution}Inner Expansion}

The inner expansion is obtained making a substitution of ${{e}^{ixt}}$  by its Taylor series expansion given by Eq.~(\ref{eq:Texp}) in the integrand of the \textcolor{black}{L\'evy-stable} distribution $I$. After this substitution, the integrals in Eq.~(\ref{eq:SD2}) and (\ref{eq:S3}) can be analytically solved. The difference between these  two equations are the truncation on the interval of integration.
  
For $\alpha\leq1$ the convergence of the series is slow, demanding a large value of order $n$ in Eq.~(\ref{eq:Texp}) to reach an acceptable similarity with the original integrand ${I}$. For this reason, the improper integral is truncated after a  small enough amplitude of $I$ is obtained.
For $\alpha>1$ the convergence occurs faster and truncation is not needed.

\subsubsection{ {For} ${0< \alpha \leq  1}$ }

The inner expansion is obtained by substituting ${{e}^{ixt}}$ in Eq.~(\ref{eq:S3}) by its Taylor expansion using Eq.~(\ref{eq:Texp}). Then:
\begin{equation}\label{eq:SD_I1}
\begin{split}
S_{i}(x;\alpha,\epsilon)= \frac{1}{\pi }\int\limits_{0}^{\tau_{2}(x,\epsilon)}{{{e}^{-{{t}^{\alpha }}}}{{e}^{ixt}}dt}  \sim  \frac{1}{\pi }\int\limits_{0}^{{{\tau}_{2}(x,\epsilon)}}{I_{n}dt}\quad as \quad n\rightarrow \infty ,
\end{split}
\end{equation}
where $I_{n}$ is given by:
\begin{equation}\label{eq:In}
\begin{split}
I_{n}(x;t,\alpha)=\sum_{k=0}^{n}e^{{-t}^{\alpha}}\frac{(ixt)^k}{k!}.
\end{split}
\end{equation}
The upper limit $\tau_{2}$ is given by the following equation:
\begin{equation}
{{\tau}_{2}}(x,\epsilon)=-\frac{\ ln (\epsilon)}{x}.
\end{equation}
\begin{figure*} [htbp]
  \centering
  \includegraphics[scale=0.47,trim=6.5cm 2cm 0.5cm 0cm]{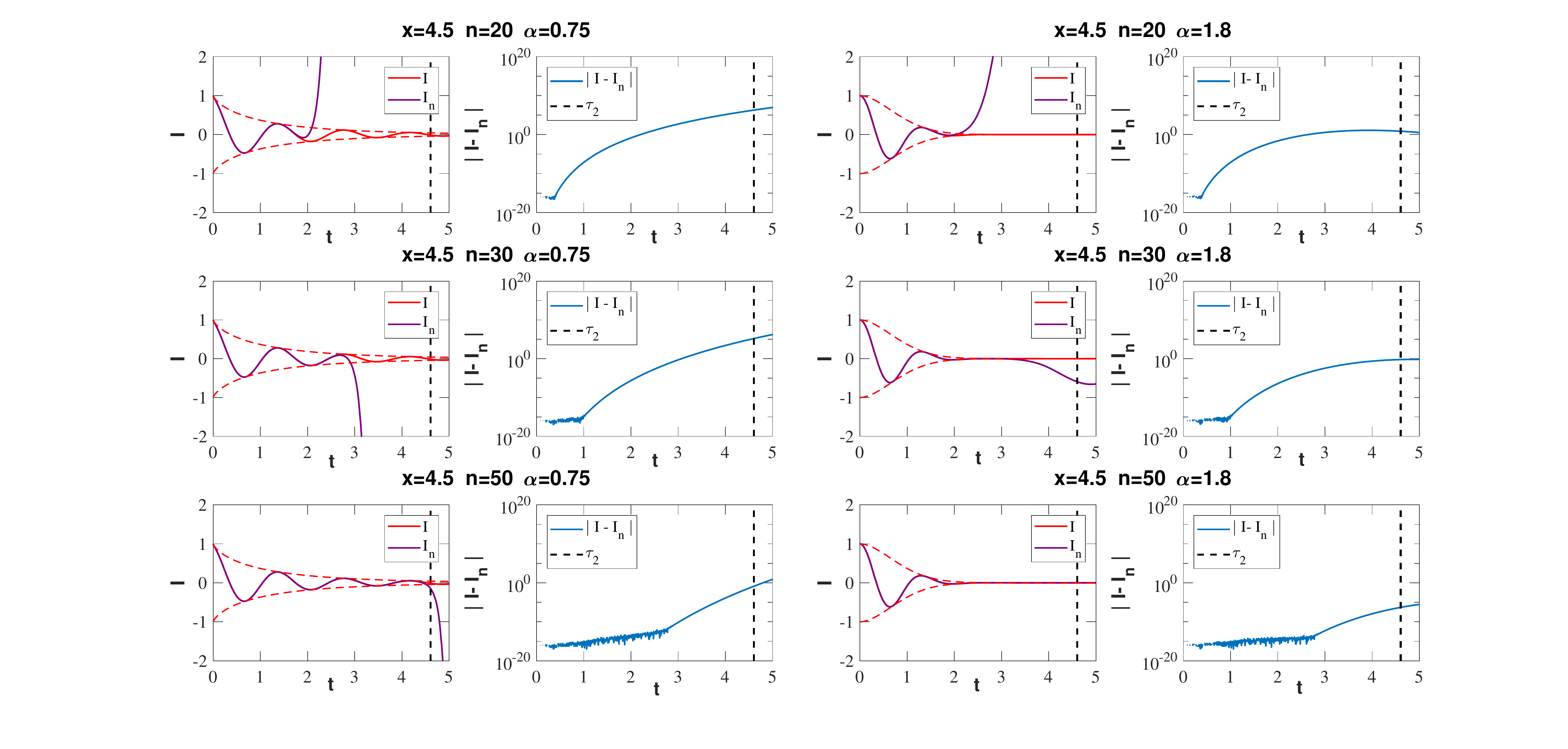}
  \caption{ \textit{
  Comparison of $I=e^{{t}^{\alpha}}e^{ixt}$ and $I_{n}=e^{{t}^{\alpha}}\sum\limits_{k=0}^{n }{\frac{{{\left( ixt \right)}^{k}}}{k!}}$ 
  of Eq.~(\ref{eq:SD_I1}) and (\ref{eq:SD_I2}), where $I_n$ is obtained by replacing $e^{ixt}$ in $I$ by its Taylor expansion.  The plots are for $\alpha = 0.75$ (left) and $\alpha =1.8$ (right). For $ \alpha \leq 1$, the truncation  is required to  ensure a cut-off when negligible quantities of $|I|$ and $|I_{n}|$ are obtained. The truncated error of the Taylor expansion is measured by using the absolute value of the difference $\left |I-I_{n}  \right |$. For $\alpha>1$, since the convergence is fast the truncation is unnecessary. For these particular examples, the integrand $I$ is evaluated at $x=4.5$ for three cases of $n=20,30,50$ with  $\epsilon=10^{-9}$.}}
  \label{fig:Fig03_Truncation}
  \end{figure*}
This truncation results from the equation ${{e}^{ix{{\tau}_{2}}}}=\epsilon$, where 
$\epsilon$ represents the tolerance that needs to be small to ensure a cut-off when negligible quantities of $|I|$ and $|I_{n}|$ are obtained. Consequently, the area under the curve of both functions are similar.
 
The convergence of $I_n$ to $I$ demands a large value of order $n$ in  Eq.~(\ref{eq:Texp}), as it can be observed in Fig~(\ref{fig:Fig03_Truncation}). This occurs because of slow decay of ${{e}^{-{{t}^{\alpha }}}}$ value for $\alpha<1$. This is the reason to evaluate the integral in the closed interval [0,${{\tau}_{2}}$],  where the original integrand $I$ and its Taylor series approximation $I_{n}$ are similar.

The integrals in Eq.~(\ref{eq:SD_I1}) can be solved without difficulty. Then, the inner expansion $s_{i}$ is given by the real part of this solution,

\begin{equation}\nonumber
s_{i}(x;\alpha,\epsilon)=Re(S_{i}(x;\alpha,\epsilon)).
\end{equation}
Consequently,
\begin{equation}\label{eq:I1}
{{s}_{i}}(x;\alpha,\epsilon)=\frac{1}{\pi \alpha }\sum\limits_{k=0}^{\infty }{\frac{{{ x }^{k}}}{k!}\gamma \left(\frac{k+1}{\alpha },{{\tau}_{2}(x;\epsilon)}^{\alpha } \right) }\cos \left( \frac{\pi k}{2} \right),
\end{equation}
where $\gamma $ represents the incomplete gamma function \citep{Abramowitz1965},
\begin{equation}\label{eq:incgamma}
\gamma(z,b)=\int_{0}^{b } {x^{z-1}}{e^{-x}}dx.
\end{equation}
Due to a computation of the incomplete gamma function $\gamma$, Eq.~(\ref{eq:I1}) was modified for numerical analysis in Matlab\footnotemark[2] \citep{didonato1986computation}. 

\footnotetext[2]{Note: Matlab defines the incomplete gamma function as  $\gamma^{*}$ 
\begin{equation}\nonumber
\gamma^{*}(b,z)=\frac{1}{\Gamma(z)}\int_{0}^{b } {x^{z-1}}{e^{-x}}dx 
\end{equation}\\where $\Gamma(z)$ is the gamma function.}

\subsubsection{{For} ${1<\alpha <2}$  }

Here it is derived the inner expansion $s_{i}$ for $\alpha>1$ from the non-truncated form of \textcolor{black}{L\'evy-stable} distribution function. This derivation is made by substituting $e^{ixt}$ in Eq.~(\ref{eq:SD2}) by  its Taylor expansion in Eq.~(\ref{eq:Texp}), then:
\begin{equation}\label{eq:SD_I2}
S_{i}(x;\alpha)= \frac{1}{\pi }\int\limits_{0}^{\infty }{{{e}^{-{{t}^{\alpha }}}}{{e}^{ixt}}dt} \sim \frac{1}{\pi }\int\limits_{0}^{\infty }{I_{n}dt} \quad as \quad n\rightarrow \infty.
\end{equation}
For  $\alpha >1 $, the convergence of  integrand $I$ and the integrand after the  substitution $I_{n}$ occurs faster than for $\alpha<1 $. This feature is observed in Figure~(\ref{fig:Fig03_Truncation}), where  an acceptable convergence between $I$ and $I_{n}$  is obtained with a small $n$ value. Consequently, the integral is evaluated without truncation or taking the limit $\epsilon\rightarrow0$ in Eq.~(\ref{eq:SD_I1}).\par 

Then, it is only considered the real part of the solution of Eq.~(\ref{eq:SD_I2}),
\begin{equation}\nonumber
s_{i}(x;\alpha)=Re(S_{i}(x;\alpha)).
\end{equation}
Consequently,
\begin{equation}\label{eq:I2}
{{s}_{i}}(x;\alpha)=\frac{1}{\pi \alpha }\sum\limits_{k=0}^{\infty }{\frac{{{ x }^{k}}}{k!}\Gamma \left(\frac{k+1}{\alpha } \right)}\cos \left( \frac{\pi k}{2} \right),
\end{equation}
where $\Gamma $ represents the gamma function \citep{Abramowitz1965},
\begin{equation}\nonumber
\Gamma(b)=\int_{0}^{\infty} {x^{b-1}}{e^{-x}}dx.
\end{equation}

 \begin{figure}[H]\nonumber
   \centering
   \includegraphics[scale=0.2,trim=2.5cm 4.cm 3cm 2cm]{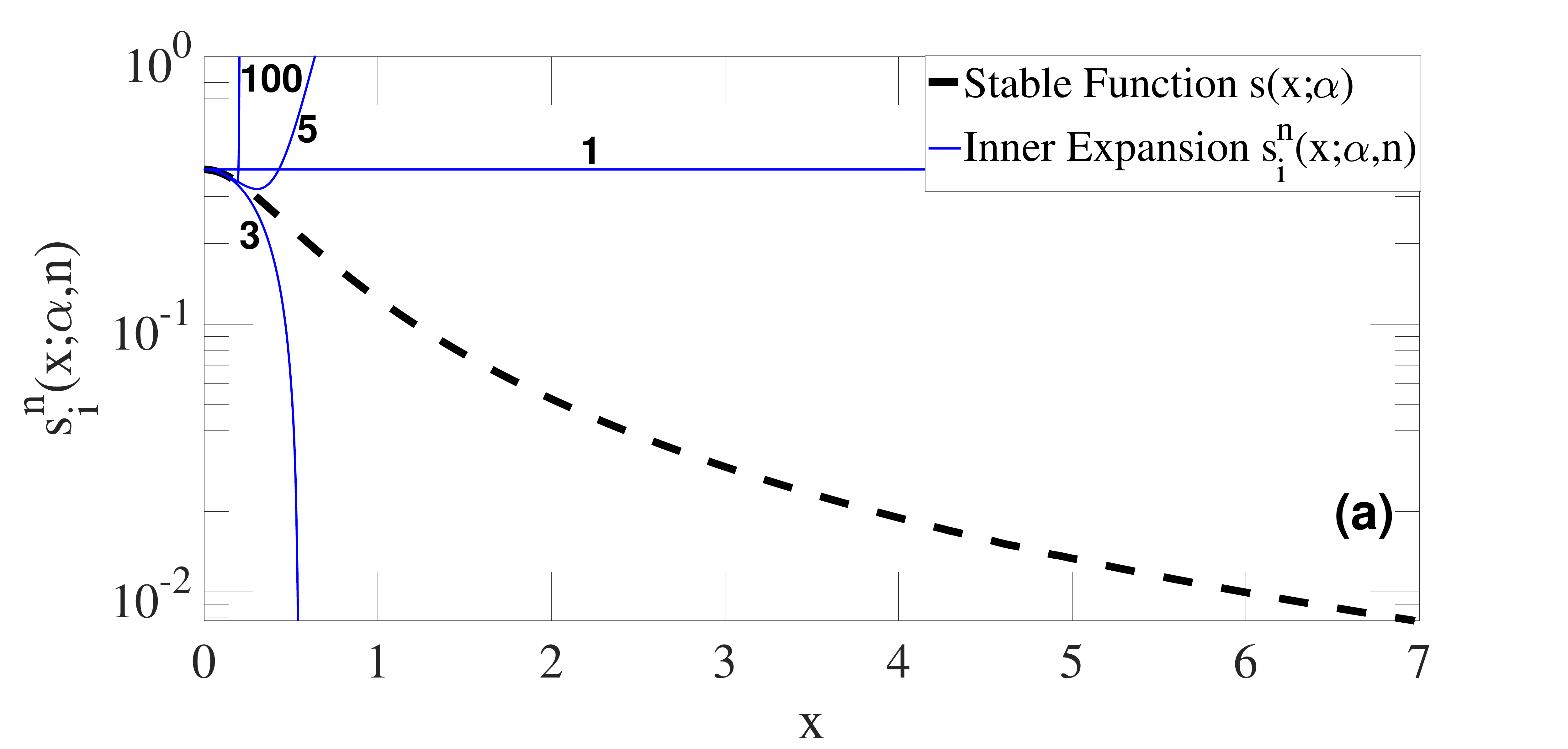}

    \end{figure}
   \begin{figure}[H]
   \centering
   \includegraphics[scale=0.2,trim=2.5cm 2cm 3cm 0cm]{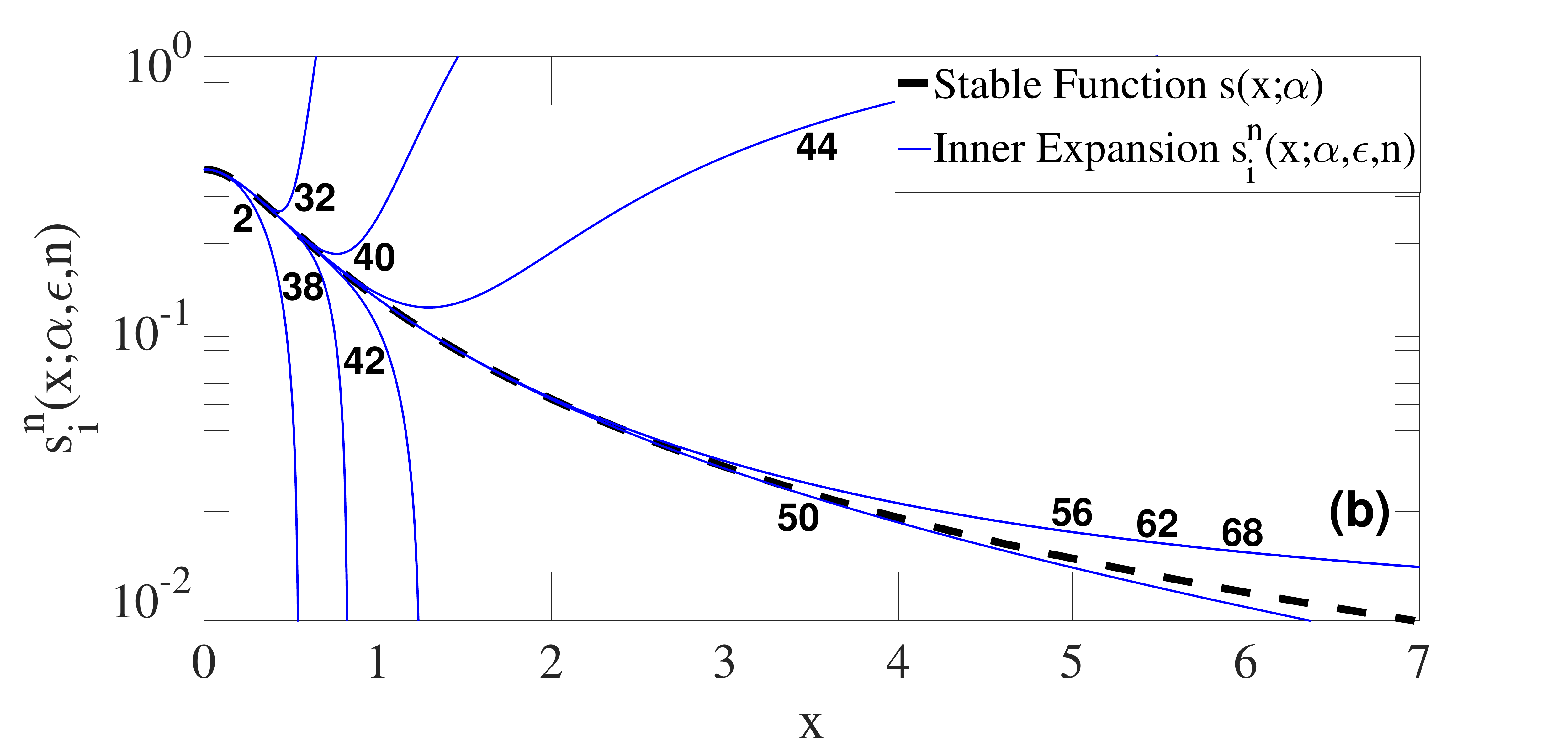}
    \caption{ \textit{Inner expansion  of the \textcolor{black}{L\'evy-stable} distribution function for $\alpha=0.75$. This is obtained by applying Taylor expansion around $t=0$. The  subfigure (a) is a non-truncated integral. The subfigure (b) is the truncated integral with tolerance $\epsilon=10^{-9}$ in Eq.~(\ref{eq:I1}), which displays a  fast convergence due to integral's truncation.}}
   \label{fig:Fig03A_Inner1}
   \end{figure}
   

  \begin{figure} [H]
  \centering
  \includegraphics[scale=0.2,trim=2.5cm 2.1cm 3cm 2cm]{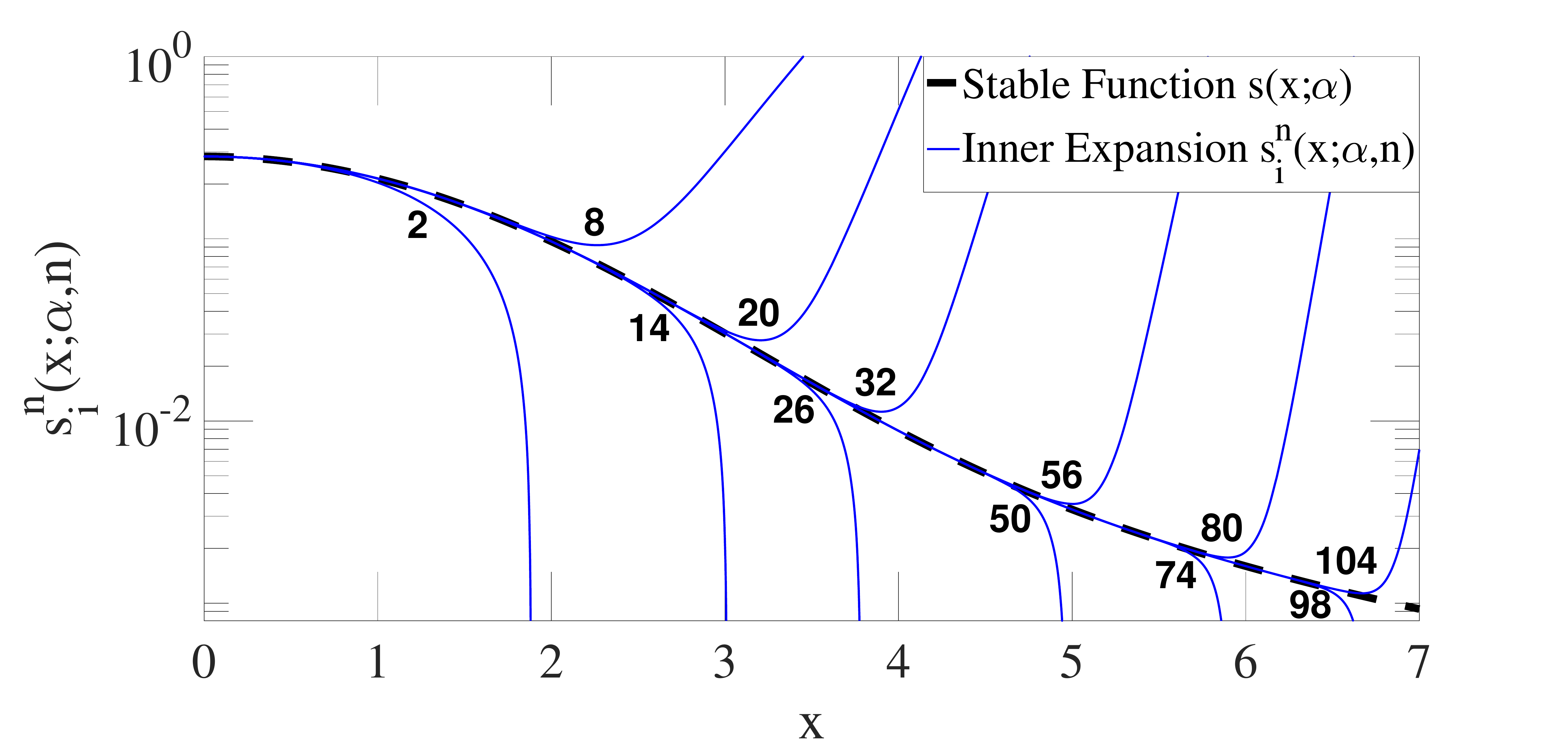}
     \caption{ \textit{Inner expansion  of the \textcolor{black}{L\'evy-stable} distribution for $\alpha=1.80$ as a result of applying a Taylor expansion in the `exponential of the phase' of the integrand. The figure illustrates the non-truncated \textcolor{black}{L\'evy-stable} solution in Eq.~(\ref{eq:I2}). This figure displays a  fast convergence so that no truncation is needed.}}
     \label{fig:Fig03B_Inner2}
     \end{figure}

Examples for $\alpha=0.75$ and $\alpha=1.80$  are shown on Figures \ref{fig:Fig03A_Inner1} and \ref{fig:Fig03B_Inner2} respectively. In Figure \ref{fig:Fig03A_Inner1}, for $\alpha\leq1$   the  truncation $\tau_{2}$ is needed, otherwise the convergence to \textcolor{black}{L\'evy-stable} distribution function will be ultraslow as $n\rightarrow\infty$. This is evident  when a comparison is made between subfigure \ref{fig:Fig03A_Inner1}a and \ref{fig:Fig03A_Inner1}b. They represent a non-truncated and truncated \textcolor{black}{L\'evy-stable} solution respectively. The subfigure \ref{fig:Fig03A_Inner1}b displays an acceptable convergence with a smaller order $n$.  In Figure \ref{fig:Fig03B_Inner2}, for $\alpha>1$  the convergence to the \textcolor{black}{L\'evy-stable} distribution function occurs faster  and no truncation is needed. For both cases the inner expansion $s_{i}$ behaves well because it converges to $s(x;\alpha)$.\par\par

\subsection{\label{OuterSolution}Outer Expansion}

The outer expansion is obtained making a substitution of the amplitude  ${e^{-t}}^\alpha$ in the integrand of the truncated \textcolor{black}{L\'evy-stable} distribution function $I$ in Eq.~(\ref{eq:S3}) by its Taylor series expansion around $t=0$. Then, the following relation is obtained:\par

\begin{equation}\label{eq:SD_O1}
\begin{split}
  S_{o}(x;\alpha,\epsilon)=
    \frac{1}{\pi }\int\limits_{0}^{\tau_{3}(\epsilon) }{{{e}^{-{{t}^{\alpha }}}}{{e}^{ixt}}dt} \sim \frac{1}{\pi } \int\limits_{0}^{{\tau_{3}(\epsilon)}}{G_{n}dt} \quad as \quad n\rightarrow \infty,
\end{split}
\end{equation}
where $G_{n}$ is given by:
\begin{equation}\label{eq:Gn}
\begin{split}
G_{n}(x;\alpha)=\sum_{k=0}^{n}{\frac{{{\left( -{{t}^{\alpha }} \right)}^{k}}}{k!}}{{e}^{ixt}}.
\end{split}
\end{equation}
The upper limit $\tau_{3}$ is given by the following equation:
 \begin{equation}
  {{\tau}_{3}}(\epsilon)={{\left[ -\ln (\epsilon) \right]}^{1/\alpha }}.
 \end{equation}
 \begin{figure*} [htbp]
  \centering
  \includegraphics[scale=0.47,trim=6.5cm 1cm 2.5cm 1.2cm]{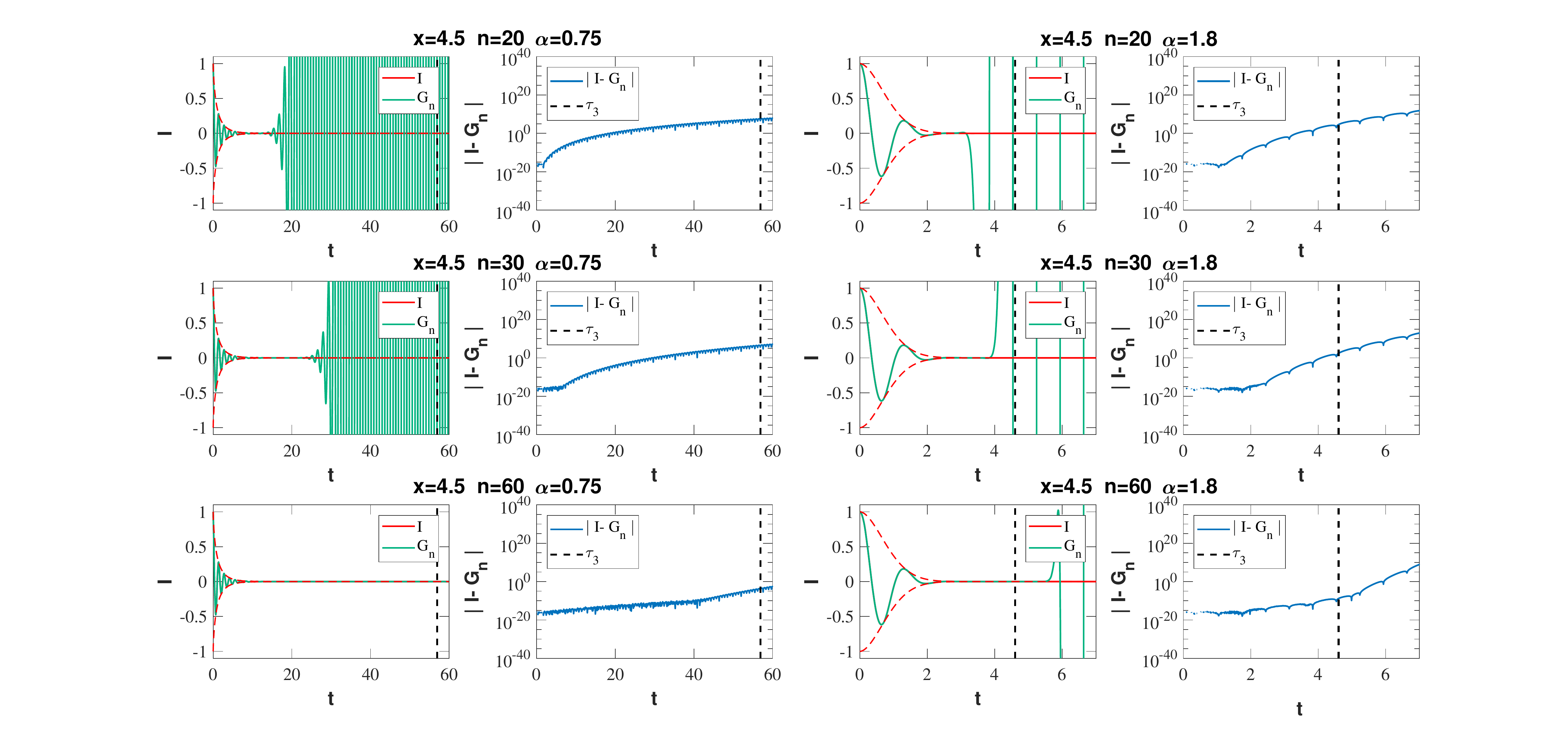}
  \caption{ \textit{Comparison of $I=e^{{t}^{\alpha}}e^{ixt}$ and $G_{n}={\frac{{{\left( -{{t}^{\alpha }} \right)}^{k}}}{k!}}{{e}^{ixt}}dt$ of Eq.~(\ref{eq:SD_O1}), where $G_n$ is the Taylor expansion of $I$ around $t=0$. The plots correspond to $\alpha = 0.75$ (left) and $\alpha = 1.8$  (right). Truncation of the integral is required for both cases. The reason of that is to reach an accurate approximation between the original integrand $I$ and the one after the series expansion $G_{n}$. The error is measured by the absolute value of the difference $\left |I-G_{n}  \right |$. For these particular examples, the integrand $I$ is evaluated at $x=4.5$ for three cases of $n=20,30,50$ with  $\epsilon=10^{-9}$.}}
  \label{fig:Fig04_Truncation2}
  \end{figure*}
This truncation is calculated from $ {{e}^{-{{\tau}_{3}}^{\alpha }}}=\epsilon$, where $\epsilon$ is defined as tolerance and represents a negligible instantaneous amplitude when $\epsilon$ is small. The truncation allows a faster convergence of  $G_{n}$ to $I$ and reduces the error of integration due to an accurate approximation on the interval $[0,\tau_{3}]$.

The original integrand $I$ and the new integrand after applying Taylor series $G_{n}$ in Eq.~(\ref{eq:SD_O1}) were evaluated in Figure~\ref{fig:Fig04_Truncation2}. Since the convergence of  $G_{n}$ to $I$ is slow, the truncation ${{\tau}_{3}}$ is considered to define the new interval of integration $[0,\tau_{3}]$.

To obtain the outer solution $s_{o}$  a change of variable after the series expansion is applied in  Eq.~(\ref{eq:SD_O1}).
The change of variable is  $-u=ixt$, so $-{du}={ix}dt$. This gives us an approximation of the form:

\begin{equation}\label{eq:O}
\begin{split}
S_{o}(x;\alpha,\epsilon)\sim
\frac{1}{\pi } \sum\limits_{k=0}^{n }{\frac{{{\left( -1 \right)}^{k}}}{k!}{{\left( \frac{-1}{ix} \right)}^{k\alpha +1}}\int\limits_{0}^{-ix{{\tau}_{3}(\epsilon)}}{{{u}^{k\alpha}}}}{{e}^{-u}}du. 
\end{split}
\end{equation}

To solve the integral, the  incomplete gamma function of imaginary argument $\gamma(v,iz)$ is used \citep{barakat1961evaluation,Abramowitz1965}. The following solution is presented by Barak as a special case of confluent hypergeometric function \citep{barakat1961evaluation},

\begin{equation}\label{eq:Chebyshev}
\begin{split}
\gamma(v,iz)&=\int_{0}^{iz}{t^{v-1}e^{-t}}dt \\ 
 \quad&\quad\\ 
 & =(iz)^{v}v^{-1}{}_{1}{{F}_{1}}(v,1+v,-iz),
\end{split}
\end{equation}
where $_{1}{{F}_{1}}(v,1+v,-iz)$ represents the Confluent Hypergeometric function. Then, comparing Eq.~(\ref{eq:O}) and (\ref{eq:Chebyshev}), we obtain the following relation between the variables, $v={k\alpha +1}$, $z=-x\tau_{3}$ and $t=u$.\par Finally, the real part of the solution is:
\begin{equation}\nonumber
s_{o}(x;\alpha,\epsilon)=Re(S_{o}(x;\alpha,\epsilon)).
\end{equation}
Consequently,
\begin{equation}\label{eq:O1}
\begin{split}
{{s}_{o}}(x;\alpha,\epsilon)=&-\frac{1}{\pi }\sum\limits_{k=1}^{\infty }\frac{{{\left( -1 \right)}^{k}}}{k!}\left( \frac{\cos \left( \pi \alpha k \right)}{k\alpha +1} \right)\quad... \\\quad... &{{(-{{\tau}_{3}(\epsilon)})}^{k\alpha +1}}{}_{1}{{F}_{1}}(k\alpha +1,k\alpha +2,ix{{\tau}_{3}(\epsilon)}).
\end{split}
\end{equation}
Figure~\ref{fig:Fig06_O1} shows the calculation of Eq.~\ref{eq:O1} for $\alpha=1.8$.	In this figure is evident that the  outer expansion $s_{o}$ converges slowly. This occurs due to computation of the confluent hypergeometric function $_{1}{{F}_{1}}$ which demands considerable computational time. The series that define the function $_{1}{{F}_{1}}$ do not have a trivial structure, this  creates numerical issues which makes the calculation computational inefficient \citep{pearson2009computation}.
The approximation in  Figure~\ref{fig:Fig06_O1} shows how the convergence  demands a large value of order $n$ to obtain an accurate approximation at the tail. The convergence resembles waves that slowly start to decrease from the tails to the peak of \textcolor{black}{L\'evy-stable} distribution.   The series  until $n=30$ does not show an acceptable approximation. Only an approximation on tails  are obtained after $n=40$. For $x\rightarrow0$, the series of $s_{o}$ converges to a specific value different of the \textcolor{black}{L\'evy-stable} distribution function. The convergence to the \textcolor{black}{L\'evy-stable} distribution function is observed only for $x\rightarrow\infty$.
	\begin{figure}[H]\nonumber
	        \centering
	        \includegraphics[scale=0.2,trim=2.5cm 2cm 3cm 1.0cm]{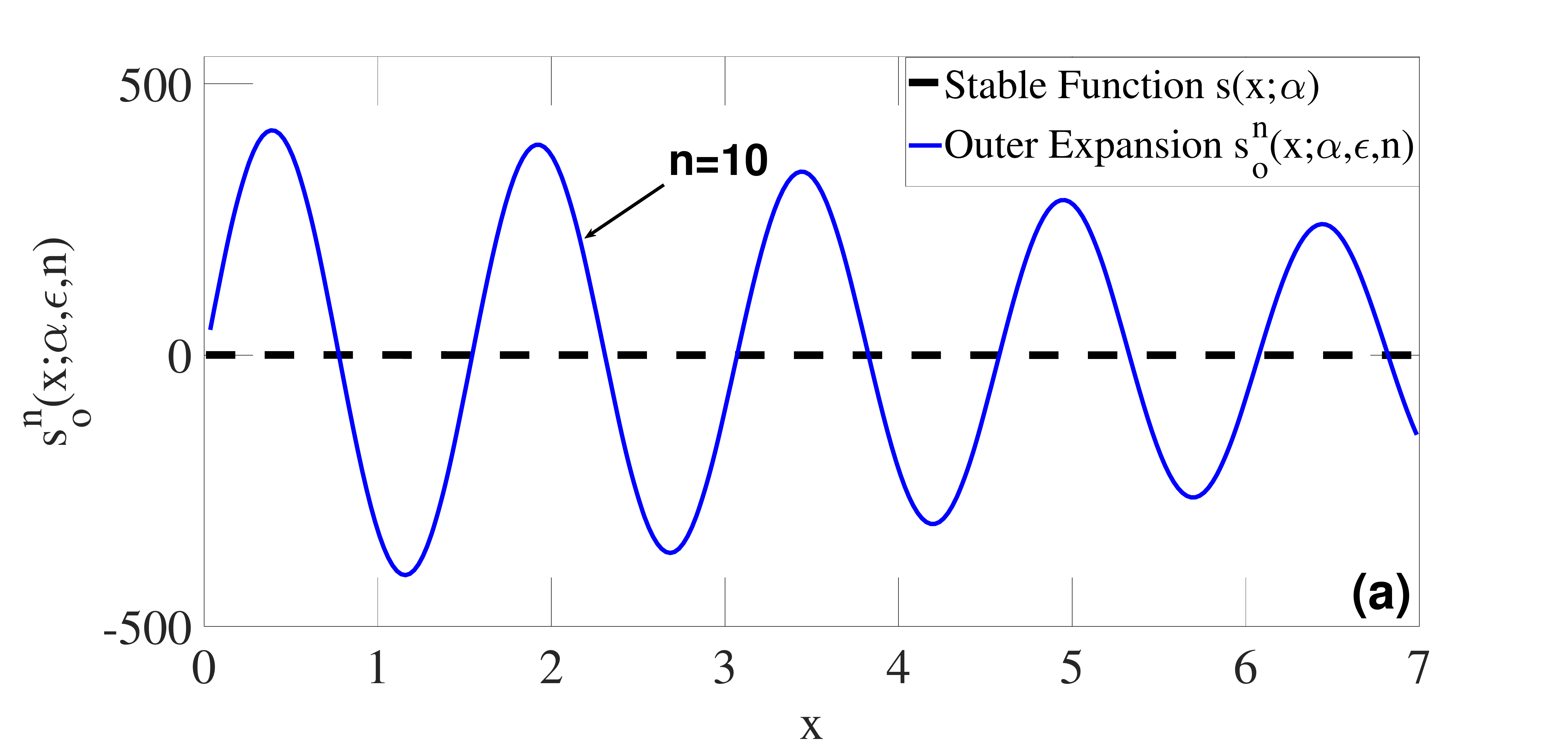}
	         \end{figure}
	         \begin{figure}[H]
	           \centering
	           \includegraphics[scale=0.2,trim=2.5cm 2cm 3cm 1cm]{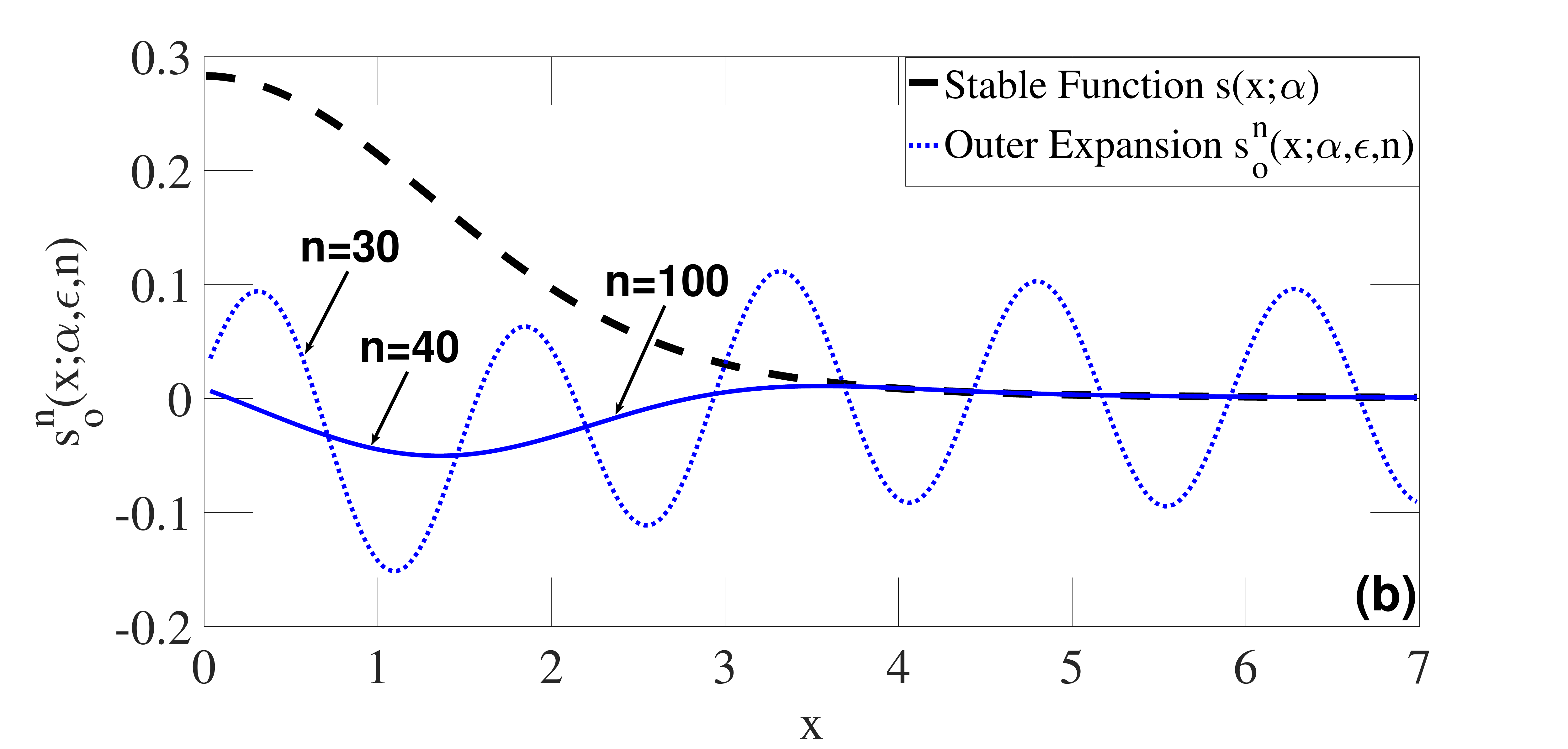}
	            \caption{  \textit{Outer expansion $s_{o}$ for $\alpha=1.80$ with tolerance $\epsilon=10^{-6}$ in Eq.~(\ref{eq:O1}). The  subfigure (a) and (b) correspond to $n=10,30,40,100$ respectively. }}
	             \label{fig:Fig06_O1}
	            \end{figure}

\subsection{\label{Outersolution2}Outer expansion by Trans-Stable distribution}

Because of the slow convergence of the outer expansion $s_{o}$ and its wave-like behaviour, an alternative approximation is obtained using the trans-stable function  $T(x;\alpha)$. As it was previously explained in section \ref{sec:trans}, the solutions of trans-stable $T(x;\alpha)$ and \textcolor{black}{L\'evy-stable} $S(x;\alpha)$ functions  are identical for ${0< \alpha \leq  1}$ and similar for ${1<\alpha <2}$ after the $x_{c}$ value. Consequently, the improper integral in Eq.~(\ref{eq:Trans-fourier and Laplace}) is used to calculate the series expansions for  $0<\alpha\leq1$ and the truncated trans-stable integral in its Laplace representation in Eq.~(\ref{eq:TS3}) for  $1<\alpha<2$.\par This outer expansion $t_{o}$ is given by the analytical solution of the trans-stable function  after applying the Taylor series  of ${e^{-(it)}}^{\alpha}$ around $t=0$ in the trans-stable integrand $\bar{I}$ using  Eq.~(\ref{eq:Texp}) and~(\ref{eq:Trans-fourier and Laplace}).   
Then, the following equation is shown:

\begin{equation}\label{eq:TS_exp}
T_{o}(x;\alpha,\epsilon)=\frac{1}{\pi }\int\limits_{0}^{\tau_{1}(x,\epsilon) }{{{e}^{-{{\left (it\right )}^{\alpha }}}}{{e}^{-xt}}idt}\sim \frac{1}{\pi }\int\limits_{0}^{\tau_{1}(x,\epsilon) }{K_{n}{dt}},
\end{equation}\\
where $K_{n}$ is given by:
\begin{equation}\label{eq:Kn}
\begin{split}
K_{n}(x)={\sum\limits_{k=0}^{n}{{\frac{{{\left( -(i{{t})^{\alpha }} \right)}^{k}}}{k!}}{{e}^{-xt}}i}}.
\end{split}
\end{equation}

Due to a slow convergence of $K_{n}$  to $\bar{I}$ the cut-off $\tau_{1}$ is applied.
The truncation $\tau_{1}$ has two different expressions. For $\alpha\leq1$, the truncation $\tau_{1}$  depends on the tolerance $\epsilon$ and for $\alpha>1$ it depends on the tolerance $\epsilon$ and $x$ values. These expressions will be explained in the following sub-sections.\par
To solve the integral in Eq.~(\ref{eq:TS_exp}), the following change of variable is applied: $xt=u$ and $xdt=du$. This leads to the following series expansion: 
\begin{equation}\label{eq:O_T}
\begin{split}
T_{o}(x;\alpha,\epsilon)\sim
\frac{1}{\pi } \sum\limits_{k=0}^{\infty }{\frac{{{\left( -1 \right)}^{k}}}{k!}{{\left( \frac{-1}{ix} \right)}^{k\alpha +1}}\int\limits_{0}^{x{{\tau}_{1}(x,\epsilon)}}{{{u}^{\left( k\alpha +1 \right)-1}}}}{{e}^{-u}}du.
\end{split}
\end{equation}
The upper limit of the integral changes from $\tau_{1}$ to $x\tau_{1}$, but still remains on the real axis.
The integral above can be solved using the incomplete gamma function defined in Eq.~(\ref{eq:incgamma}). Then, the real part of the result is obtained,

\begin{equation}\nonumber
t_{o}(x;\alpha,\epsilon)=Re(T_{o}(x;\alpha,\epsilon)).
\end{equation}
Consequently,
\begin{equation}\label{eq:O2a} 
\begin{split}
{{t}_{o}}(x;\alpha,\epsilon)=&\\-\frac{1}{\pi }&\sum\limits_{k=1}^{\infty }{\frac{{{\left( -1 \right)}^{k}}}{k!}{{\left( \frac{1}{x} \right)}^{k\alpha +1}}\sin \left( \frac{\pi \alpha k}{2} \right)\gamma \left( k\alpha +1,x{\tau}_{1}(x,\epsilon) \right)}.
\end{split}
\end{equation}
The determination of $\tau_{1}$ for $\alpha\leq1$ and $\alpha>1$ is presented in the following subsections.

\subsubsection{{For} ${0< \alpha \leq  1}$  }
For $\alpha\leq1$, the cut-off $\tau_{1}$ in Eq.~(\ref{eq:O2a}) is given by the following equation: 
  
  \begin{equation}\label{eq:O2aux}
   {{\tau_{1}(\epsilon)}}={{\left[ -\ln (\epsilon) \right]}^{1/\alpha }}\quad for \quad\alpha\leq1.
  \end{equation}
This truncation is obtained from ${{e}^{-{{\tau_{1}}}^{\alpha }}}=\epsilon$, where  the tolerance $\epsilon$  represents a negligible instantaneous amplitude for the integrands in Eq.~(\ref{eq:TS_exp}).\par

\subsubsection{{For} ${1<\alpha <2}$}
The truncation $\tau_{1}$ in Eq.~(\ref{eq:O2a}) for $1<\alpha <2$ was already obtained 
in subsection \ref{sec:trans}-2 and defined by Eq.~(\ref{eq:TS2a}) as:
\begin{equation}\nonumber
\tau_{1}(x,\epsilon) =\left\{\begin{matrix}
t_{c}(\epsilon)\quad if\quad x>x_{c},\\ 
{x}/{\alpha}\quad if\quad x<x_{c},
\end{matrix}\right.\quad for \quad\alpha>1,
\end{equation}
where $t_{c}$ and $x_{c}$ were defined by Eq.(\ref{eq:TS2b}).
As  indicated in subsection \ref{sec:trans}-2, the value of $\tau_{1}$ is used to minimize the truncation error and at the same time to make the domain of integration as small as possible.\par
The outer expansion by the trans-stable function  converges to the original trans-stable function.
Examples are shown in Figure \ref{fig:Fig04A_Outer1}  for $\alpha\leq1$ and Figure \ref{fig:Fig04B_Outer2} for $\alpha>1$. Note that in both cases the truncation $\tau_{1}$ allows a faster and more accurate convergence to the real part of the trans-stable distribution $t(x;\alpha)$. Consequently the outer solution $t_{o}$ shows an identical solution as $s(x;\alpha)$ for $\alpha\leq1$ and the same asymptotic behaviour for $\alpha>1$. For a smaller $\epsilon$ the convergence of these outer expansions to the trans-stable function will occur faster. Also in Figure \ref{fig:Fig04A_Outer1} the non-truncated trans-stable expansion is shown as an expansion that converges extremely slowly requiring a higher order $n$ than truncated trans-stable expansion to obtain an acceptable convergence. In Figure \ref{fig:Fig04B_Outer2} the non-truncated trans-stable expansion does not converge to trans-stable function at all.

 \begin{figure}[htbp]\nonumber
  \centering
  \includegraphics[scale=0.2,trim=2.5cm 6cm 3cm 5.5cm]{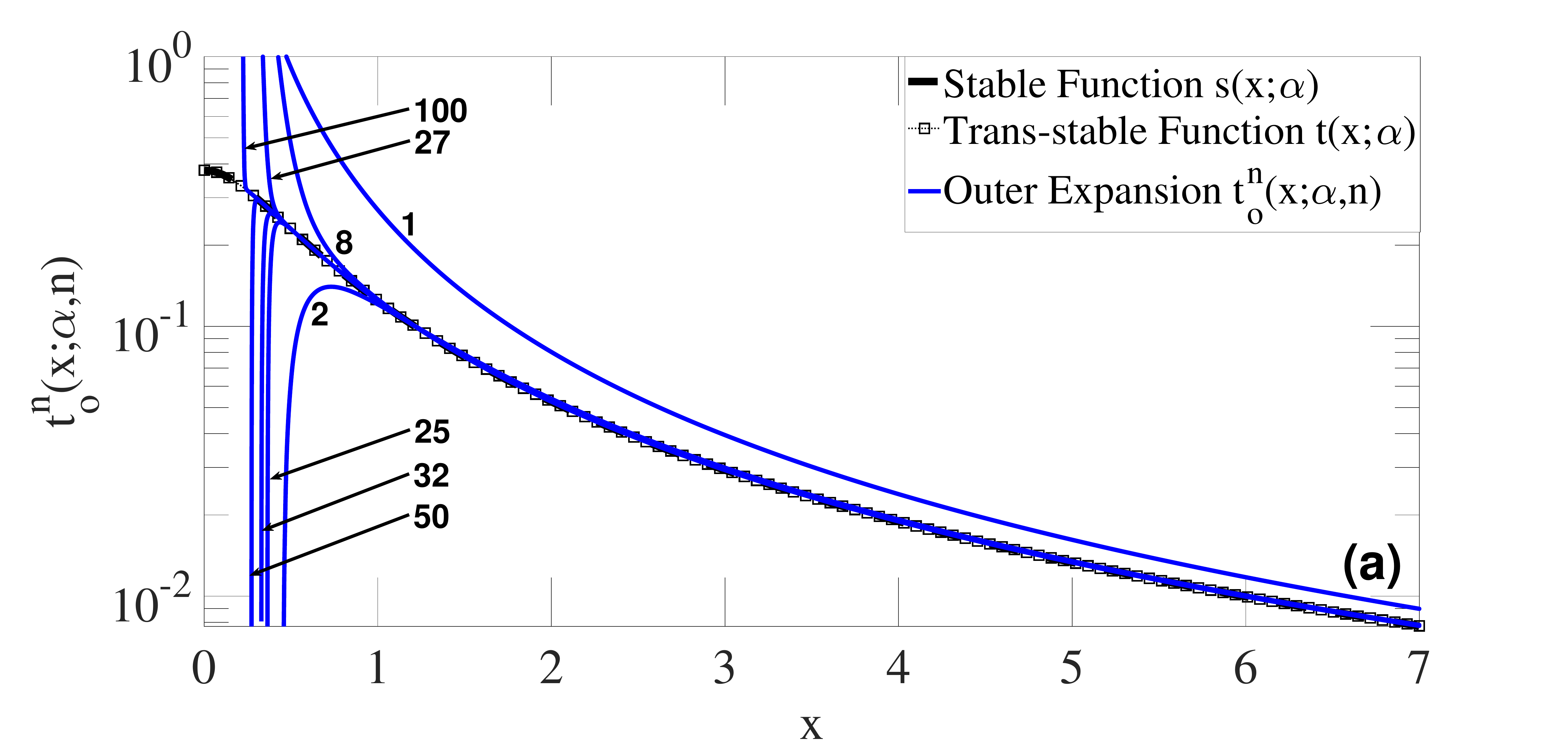}

  \end{figure}
  \begin{figure}[htbp]
   \centering
   \includegraphics[scale=0.2,trim=2.5cm 2.2cm 3cm 6cm]{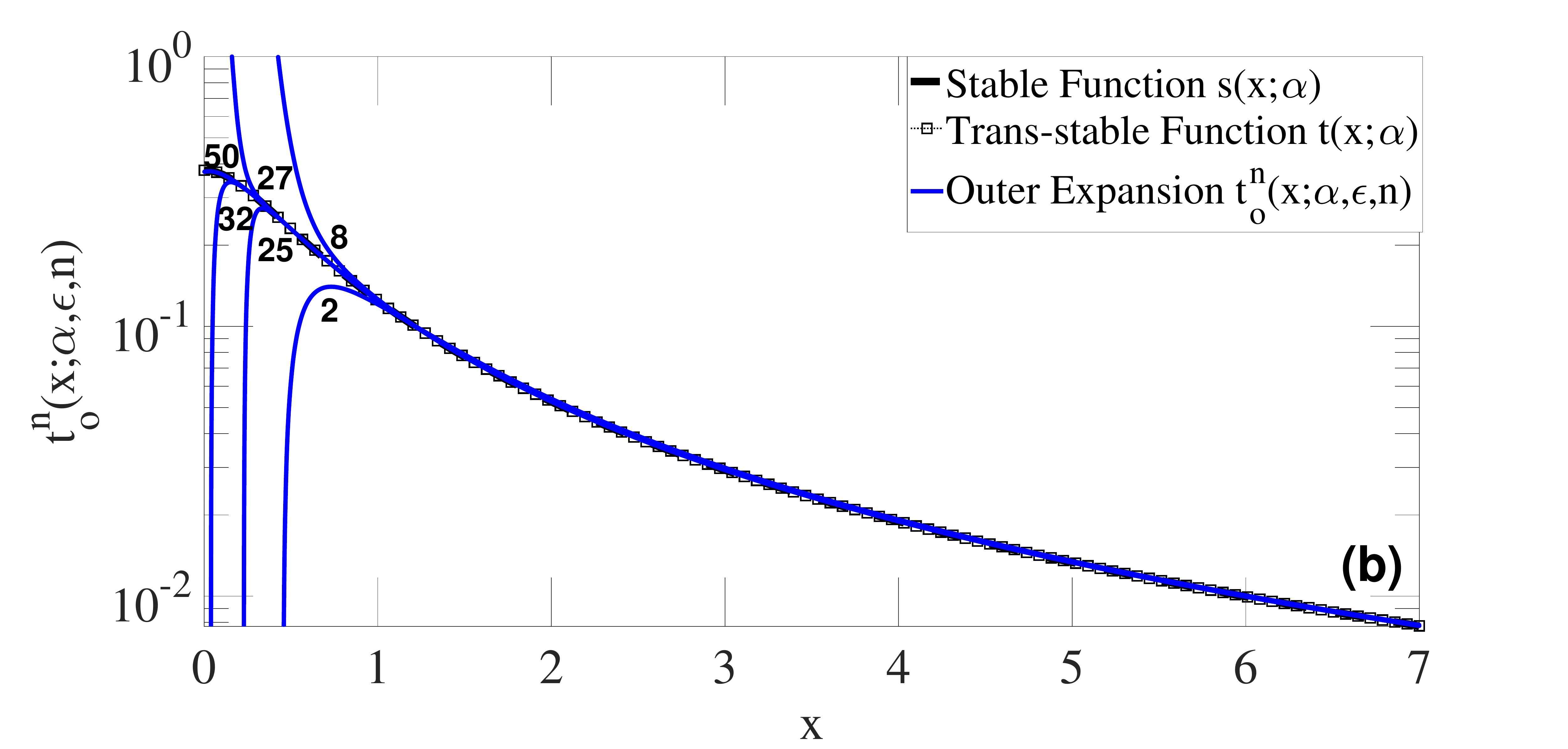}
    \caption{ \textit{ \textit{{Outer expansion of the trans-stable function for $\alpha=0.75$. This result is obtained from the Taylor expansion of the integrand around $t=0$ in Eq.~(\ref{eq:O2a}) and (\ref{eq:O2aux}). The subfigure (a) is the non-truncated integral that shows slow convergence. The subfigure (b) corresponds to truncated integral with tolerance $\epsilon=10^{-6}$. The subfigure (b) displays a faster convergence to the trans-stable function as a result of the truncation of the integral.}}}}
   \label{fig:Fig04A_Outer1}
   \end{figure}
   \pagebreak
\begin{figure}[H]\nonumber
 \centering
 \includegraphics[scale=0.2,trim=2.5cm 3cm 3cm 1.5cm]{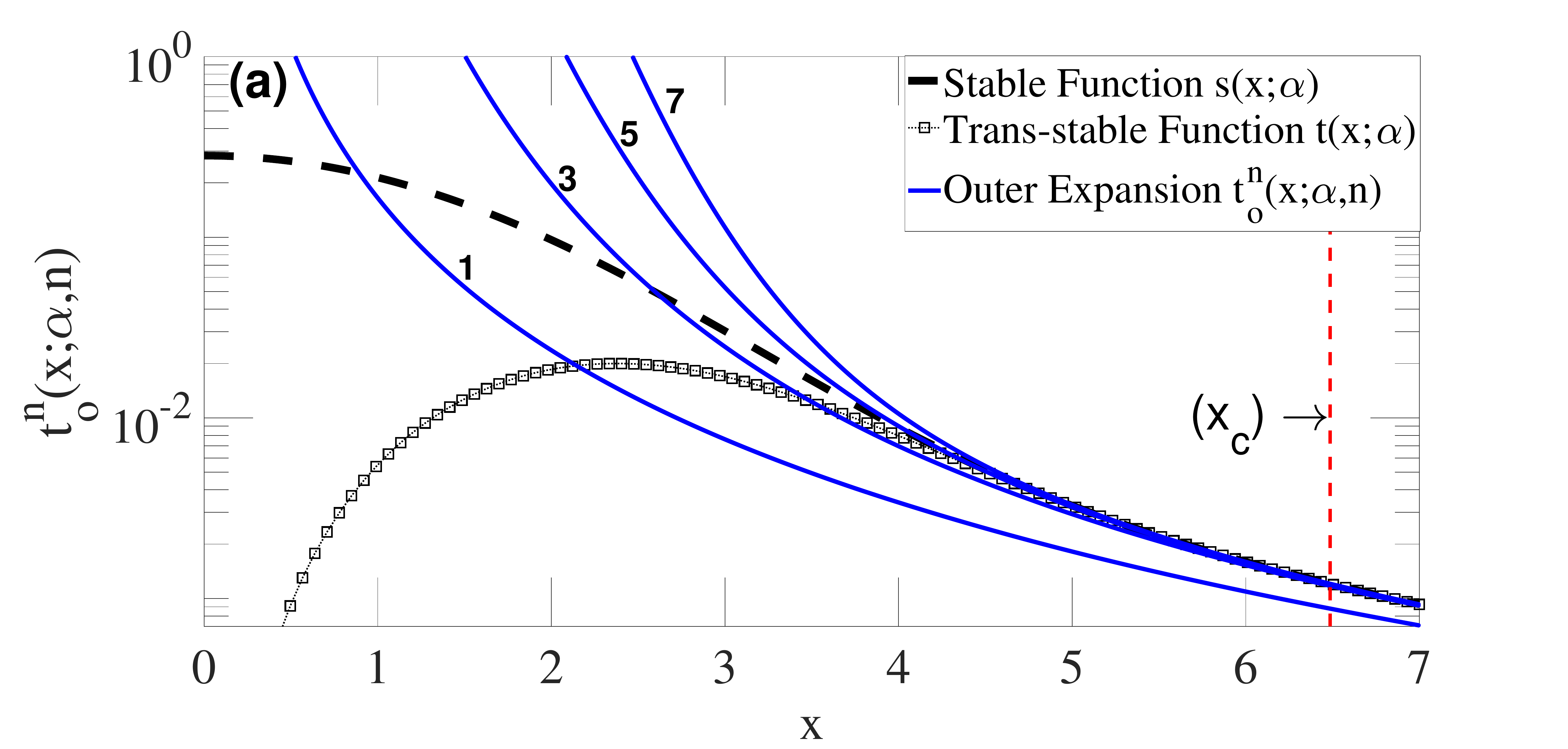}
 \end{figure}
 \begin{figure}[H]
  \centering
  \includegraphics[scale=0.2,trim=2.5cm 0.8cm 3cm 1.5cm]{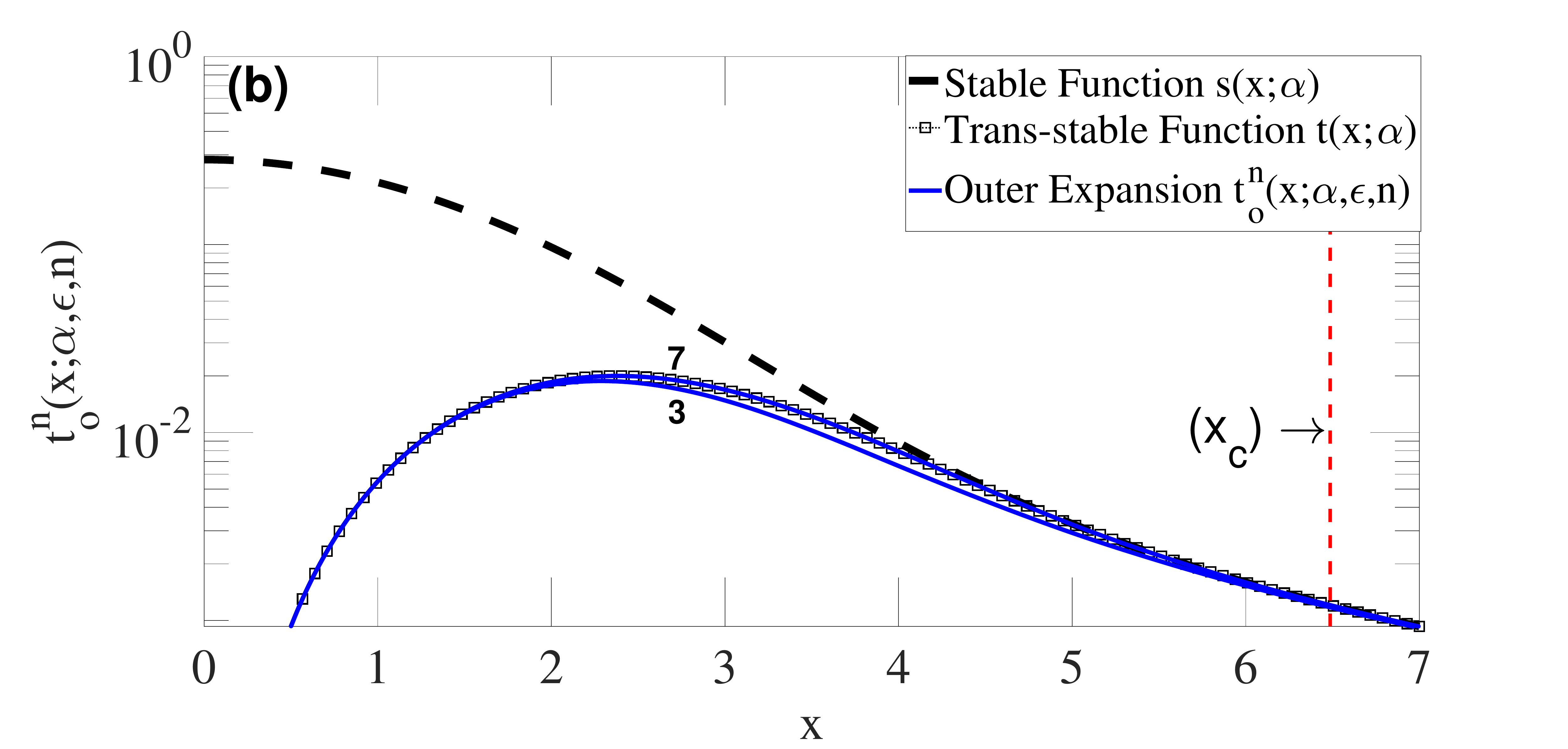}
   \caption{ \textit{{Outer expansion of the trans-stable for $\alpha=1.80$ as a result of applying Taylor expansion of the integrand around $t=0$  in Eq.~(\ref{eq:TS2a}) and (\ref{eq:O2a}). The subfigure (a) shows that the non-truncated integral does not converge to the trans-stable function. The subfigure (b) corresponds to the truncated integral with tolerance $\epsilon=10^{-6}$. The subfigure (b) displays a fast convergence as a result of the truncation of the integral.}}}
  \label{fig:Fig04B_Outer2}
  \end{figure}

\section{\label{sec:Uniform} Uniform solution}
The uniform solution is presented as the combination of the inner solution and the outer solution to construct an approximation valid for all $x \in [-\infty,\infty]$. To construct the uniform solution  an asymptotic matching method based on boundary-layer theory is applied \cite{Benderorszag1999,roos2008robust}. This method is based on superposing the inner and outer solution and subtracting the overlap between them, 
\begin{equation}\label{eq:yunif}
{{s}_{u}}(x)={{y}_{out}}(x)+{{y}_{in}}(x)-{{y}_{overlap}}(x).
\end{equation}
The overlap is defined as the limit of the rightmost edge of $y_{in}$ and the leftmost edge of $y_{out}$,\par   
\begin{equation}\label{eq:yover}
y_{overlap}=\lim_{x\rightarrow 0}y_{out}=\lim_{x\rightarrow \infty}y_{in}.
\end{equation}
For this case, our proposed uniform solution $s_{u}$  is constructed based on our inner expansion $s_{i}$ and our outer expansion $t_{o}$. These previous solutions were already defined in section \ref{sec:Asymptotic}.

For a better understanding of our uniform solution $s_{u}$, two sub-sections A and B are presented. Sub-section A contains a summary of inner and outer expansions previously obtained. In sub-section B the steps taken to obtain  $s_{u}$ are explained.

 \begin{table*}[t]
   \centering
 \begin{tabular}{|>{\em} M{4cm}|M{7.4cm}|M{7cm}|}
 
  \hline 
  
 Range of $\alpha$& \[{0< \alpha \leq  1}\] & \[{1<\alpha <2}\]  \\ 
 \hline
 \parbox{0pt}{\rule{0pt}{5ex+\baselineskip}} Normalized \textcolor{black}{L\'evy-stable} distribution  $({{s}})$  &  
  \multicolumn{2}{c|}{$s(x;\alpha)= \frac{1}{\pi }\int\limits_{0}^{\infty }{{{e}^{-t}}^{^{\alpha }}\cos (tx)dt} 
  $  }
  \\ 
  \hline
   \parbox{0pt}{\rule{0pt}{5ex+\baselineskip}} Normalized trans-stable distribution  $({{t}})$  &  
    \multicolumn{2}{c|}{$t(x;\alpha)= \frac{1}{\pi }\int\limits_{0}^{\infty }{{{e}^{-{{t}^{\alpha }}\cos \left( \frac{\pi \alpha }{2} \right)-xt}}{\sin(t^{\alpha} \sin(\frac{\pi\alpha}{2}))}dt}
    $  }
    \\ 
 \hline
 Inner expansion $({{s}_{i}^{n}})$   & \[ s_{i}^{n}(x;\alpha,\epsilon)=\frac{1}{\pi \alpha }\sum\limits_{k=0}^{n}{\frac{{{ x }^{k}}}{k!}\gamma \left(\frac{k+1}{\alpha },{{\tau}_{2}}^{\alpha } \right) }\cos \left( \frac{\pi k}{2} \right)\] \[{{\tau}_{2}}=\frac{-\ln (\epsilon)}{x}\] & \[s_{i}^{n}(x;\alpha)=\frac{1}{\pi \alpha }\sum\limits_{k=0}^{n}{\frac{{{ x }^{k}}}{k!}\Gamma \left(\frac{k+1}{\alpha } \right) }\cos \left( \frac{\pi k}{2} \right)\] \\ 
 \hline
 
 \parbox{0pt}{\rule{0pt}{10ex+\baselineskip}} Outer expansion $({s}_{o}^{n})$   &  
  \multicolumn{2}{c|}{
  
    $\begin{aligned}[t]
    {s}_{o}^{n}(x;\alpha,\epsilon)= -\frac{1}{\pi }\sum\limits_{k=1}^{n}\frac{{{\left( -1 \right)}^{k}}}{k!}&\left( \frac{\cos \left( \pi \alpha k \right)}{k\alpha +1} \right)\quad{{(-{{\tau}_{3}})}^{k\alpha +1}}_{1}{{F}_{1}}(k\alpha +1,k\alpha +2,ix{{\tau}_{3}})
     \\& {{\tau}_{3}}={{\left[ -\ln (\epsilon) \right]}^{1/\alpha }}\\ \quad
  \end{aligned}$}
  \\ 
 \hline
         
\parbox{0pt}{\rule{0pt}{10ex+\baselineskip}} {Outer solution \footnotemark[3] $({t}_{o}^{n})$ } %
  &  
  \multicolumn{2}{c|}{
  
    $ \begin{aligned}[t]
  {t}_{o}^{n}(x;\alpha,\epsilon)=-\frac{1}{\pi }\sum\limits_{k=1}^{n}{\frac{{{\left( -1 \right)}^{k}}}{k!}{{\left( \frac{1}{x} \right)}^{k\alpha +1}}\gamma \left( k\alpha +1,x{\tau}_{1} \right)\sin \left( \frac{\pi \alpha k}{2} \right)}
        \end{aligned}
  $}

                   \\
                   \cline{2-3}
                   \parbox{0pt}{\rule{0pt}{-6ex+\baselineskip}}{} 
                    & \[{{\tau}_{1}}={{\left( -\ln (\epsilon) \right)}^{1/\alpha }}\]  & \[\tau_{1} =\left\{\begin{matrix}
                           t_{c} \quad if\quad x>x_{c}\\ 
                           {x}/{\alpha}\quad if\quad x<x_{c}
                           \end{matrix}\right.\] \\ 
                   \hline
                    \parbox{0pt}{\rule{0pt}{+5ex+\baselineskip}}  Complete and incomplete gamma functions $(\Gamma)$ $\&$ $(\gamma)$
                    &  
                     \multicolumn{2}{c|}{       
               $ \begin{aligned}[t]
                                   \Gamma(z)=\int_{0}^{\infty } {x^{z-1}}{e^{-x}}dx \quad
                                   \gamma(z,b)=\int_{0}^{b } {x^{z-1}}{e^{-x}}dx \\ \quad
                                        \end{aligned}
                                 $}   
                \\ \hline
   \end{tabular}
   \caption{Summary of Inner and Outer Solutions}
   \label{tab:1}
 \end{table*}

\footnotetext[3] {Note: Refer to equation Eq.~(\ref{eq:TS2a}) to obtain $t_{c}$ and $x_{c}$ value for $1<\alpha<2$}
\subsection{\label{sec:Summary}Summary of inner and outer expansions}
 Table \ref{tab:1} contains the normalized \textcolor{black}{L\'evy-stable} and trans-stable distribution  and the summary of previous results obtained from \textcolor{black}{L\'evy-stable} and trans-stable functions by applying Taylor expansions. The series refers to one inner expansion $s_{i}$ and two outer expansions $s_{o}$ and $t_{o}$.\par 
For the inner expansion $s_i$, the solution for $\alpha\leq1$ corresponds to a truncated \textcolor{black}{L\'evy-stable} solution which allows a faster convergence. For $\alpha>1$ the series is obtained from the non-truncated \textcolor{black}{L\'evy-stable} solution. The only difference between them is the use of the incomplete gamma function $\gamma$ in the solution for $\alpha\leq1$, where $\Gamma(z) =\lim_{b\rightarrow\infty }\gamma (z,b)$. Consequently, for both cases the  truncated series can provide a good approximation. However, in the case of $\alpha\leq1$ we must take the limit as:

\[s(x;\alpha)=\lim_{\epsilon\rightarrow 0}\left ( \lim_{n\rightarrow \infty} s_{i}^{n}(x;\alpha,\epsilon) \right )\quad for\quad x<\infty.\]
In general  the order how we apply the limits cannot be exchanged. However, in the case  of $\alpha>1$ the order of the limits does not affect the convergence. Taking a small value of $\epsilon$ ensures a faster convergence.\par
For the outer expansion two expressions were derived. The first outer expansion $s_{o}$ is obtained by performing the Taylor expansion around $t=0$ on the truncated \textcolor{black}{L\'evy-stable} distribution. This solution displays a slow convergence for $n\rightarrow\infty$. The second outer expansion $t_{o}$ is obtained by applying the Taylor expansion on the truncated  trans-stable function for $x\rightarrow\infty$. The truncation of $t_{o}$ depends on $\alpha$ and there are two different cases. For $\alpha\leq1$ it converges to the exact solution of $s(x;\alpha)$  and for $\alpha>1$ it converges to the same solution at the tails of $s(x;\alpha)$. To guarantee convergence, we need to take the limit as,

\[t(x;\alpha)=\lim_{\epsilon\rightarrow 0}\left ( \lim_{n\rightarrow \infty} t_{o}^{n}(x;\alpha,\epsilon) \right )\quad for\quad x>0.\]
 
Exchanging the order of the limits will affect the convergence. The outer expansion that will be used is $t_{o}$, because it displays a faster convergence and it does not exhibit wavelike behaviour.

\subsection{\label{sec:Steps}Steps to obtain the uniform solution}
To obtain the uniform solution $s_{u}$ the condition in Eq.~(\ref{eq:yover}) needs to be satisfied. The inner expansion $s_i$ and the outer expansion $t_{o}$  have to be multiplied with an appropriate coefficient $A(x)$ to obtain the asymptotic solutions with a common matching value  $y_m$. These operations will allow us to obtain $y_{out}$ and $y_{in}$. Consequently, Eq.~(\ref{eq:yunif}) will be applied to obtain the closed-form solution of the \textcolor{black}{L\'evy-stable} distribution function.\par
Below the steps are explained to obtain the location of the matching between the inner and the outer solutions $({x}_{m},{y}_{m})$, the coefficient $A(x)$, and the uniform solution $s_{u}$. 

\subsubsection{{Finding inner and outer limit }${(x_{m},y_{m})}$ }
 Considering $s_{i}$ and $t_{o}$ as good approximations to the \textcolor{black}{L\'evy-stable} distribution function,  we must require that the inner and the outer expansions will be close enough before matching them \cite{holmes2012introduction}.  
  Consequently, the point where the matching between ${{s}_{i}}$ and ${{t}_{o}}$ takes place is $({x}_{m},{y}_{m})$ and it represents the location where the minimal vertical distance between the inner ${{s}_{i}}$ and the outer solution ${{t}_{o}}$ occurs.\par 
 The distance function between ${{s}_{i}}$ and $s$  is defined as  $\delta_{i}$ and the distance function between ${{t}_{o}}$ and $s$ is $\delta_{o}$.
 Consequently, $({x}_{m},{y}_{m})$ is the point where the Pythagorean addition of these distances is minimal Eq.~(\ref{eq:xm}):

\[{{\delta }_{i}}^{2}(x;\alpha,\epsilon)={{\left( s(x;\alpha)-{{s}_{i}}(x;\alpha,\epsilon) \right)}^{2}},\]
\[{{\delta }_{o}}^{2}(x;\alpha,\epsilon)={{\left( s(x;\alpha)-{{t}_{o}}(x;\alpha,\epsilon) \right)}^{2}},\]
\[{{\delta }^{2}}(x;\alpha,\epsilon)={{\delta }_{o}}^{2}(x;\alpha,\epsilon)+{{\delta }_{i}}^{2}(x;\alpha,\epsilon),\]

\begin{equation}\label{eq:xm}
{{\left. \frac{d\left( {{\delta }^{2}}(x;\alpha,\epsilon) \right)}{dx} \right|}_{{{x}_{m}}}}=0.
\end{equation}
The $x_{m}$ value is obtained from the previous equation. Then, $y_{m}$ is defined by the equidistant point between both functions,
\begin{equation}\label{eq:ym}
{{y}_{m}}=\frac{{{s}_{i}}({{x}_{m}})+{{t}_{o}}({{x}_{m}})}{2}.
\end{equation}

\subsubsection{{Defining the inner and the outer solutions  ${y_{in}}$ and ${y_{out}}$ }}
To obtain the uniform solution ${{s}_{u}}$, the asymptotic matching method based on boundary layer theorem \cite{Benderorszag1999} is applied. 
Consequently, the inner solution $y_{in}$ and the outer solution $y_{out}$ must have a matching asymptotic behaviour. More precisely, the limit of the outer solution $y_{out}$ when $x\rightarrow0$ should correspond to the limit of the inner solution $y_{in}$ when $x\rightarrow\infty$. 
 To obtain $y_{in}$ and $y_{out}$ solutions, the series expansions  $s_{i}$ and $t_{o}$ are multiplied by an appropriate coefficients to meet the requirements of matching asymptotic expansions, so the $y_{in}$ and $y_{out}$ are defined as follows:

\begin{equation}\label{eq:yin}
 y_{in}(x;\alpha,\epsilon,\mu)=({{s}_{i}}(x)-{{y}_{m}})\left( 1-A(x;\mu) \right)+{{y}_{m}},  
\end{equation}

\begin{equation}\label{eq:yout}
y_{out}(x;\alpha,\epsilon,\mu)=({{t}_{o}}(x)-{{y}_{m}})A(x;\mu)+{{y}_{m}},
\end{equation}
where the overlapping factor $A(x)$ is defined as:

\begin{equation}
A(x;\mu)=\frac{1}{2}\left( 1+\tanh \left( \frac{x-{{x}_{m}}}{\mu } \right) \right).
\end{equation}
The $A(x;\mu)$ is used to smooth $s_{i}$ and $t_{o}$ and provides them with a symmetric overlap section around $x_{m}$ and gives $y_{in}$ and $y_{out}$ an asymptotic behaviour. The variable $\mu$ determines the width of the overlap between $y_{in}$ and $y_{out}$.\par
It is easy to see that Eq.~(\ref{eq:yin}) and Eq.~(\ref{eq:yout}) satisfy Eq.~(\ref{eq:yover}), where the limits of $y_{out}$ and $y_{in}$ converge to a constant value $y_{m}$.

\subsubsection{Defining the Uniform Solution ${{s}_{u}}$}
The inner solution $y_{in}$ Eq.~(\ref{eq:yin}) and  the outer solution $y_{out}$ in Eq.~(\ref{eq:yout}) were defined to fulfill the requirements for matching asymptotic expansions. Then, Eq.(\ref{eq:yunif}) is applied to obtain the uniform solution $s_{u}$,
\begin{equation}\label{eq:Su}
\begin{split}
{{s}_{u}^{n_{i},n_{o}}(x;\alpha,\epsilon,\mu)}=&\frac{{{t}_{o}^{n_{o}}}(x;\alpha,\epsilon)}{2}+\frac{{{s}_{i}^{n_{i}}}(x;\alpha,\epsilon)}{2}....\\....+&\tanh\left( \frac{x-{{x}_{m}}}{\mu } \right)\left( \frac{{{t}_{o}^{n_{o}}}(x;\alpha,\epsilon)}{2}-\frac{{{s}_{i}^{n_{i}}}(x;\alpha,\epsilon)}{2} \right). 
\end{split}
\end{equation}

\subsubsection{{Find the best ${{s}_{u}}$ by choosing the most appropriate $\mu $}}
The width of the overlap between $y_{in}$ and $y_{out}$ can be optimized to obtain the closest solution $s_{u}$ of the \textcolor{black}{L\'evy-stable} distribution function. The most appropriate value of $\mu$ needs to be obtained for each particular value of $\alpha$. For that, the least square method will be applied between the original $s(x;\alpha)$ and the new closest solution ${{s}_{u}}(x;\alpha,\epsilon,\mu)$. Applying Eq.~(\ref{eq:SD_1}) and (\ref{eq:Su}) the following equation is obtained:

\begin{equation}\label{eq:opt_mu}
L(\mu)=\sum_{i=1}^{N}{{{\left( {{s}_{u}}(x_{i};\alpha,\epsilon,\mu)-s(x_{i};\alpha) \right)}^{2}}},
\end{equation}
where the $N$ value represents the length of the sample used to minimize L.\\

\begin{figure}[H]
 \centering
 \includegraphics[scale=0.2,trim=2.5cm 2.0cm 3cm 0cm]{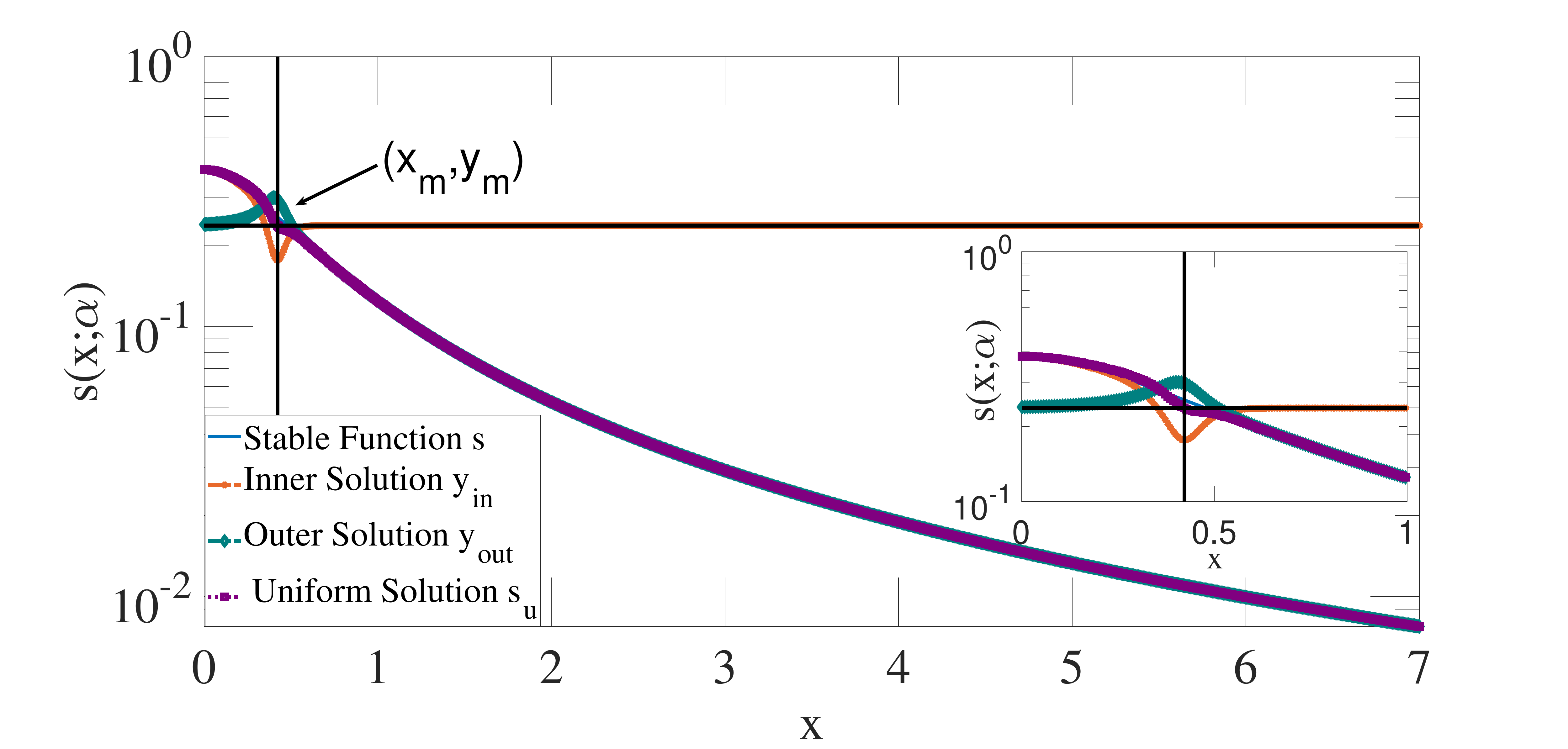}
  \caption{ \textcolor{black}{\textit{Uniform solution $s_{u}$ for $\alpha=0.75$ as a result of joining the inner solution $y_{in}$ with the outer solution $y_{out}$. The tolerance $\epsilon=10^{-6}$, $\mu=0.052$ and $n_{i}=6$ and $n_{o}=17$.}}}
 \label{fig:Fig05A_Uniform1}
 \end{figure}

\begin{figure}[H]
  \centering
 \includegraphics[scale=0.2,trim=2.5cm 1.0cm 3cm 2.0cm]{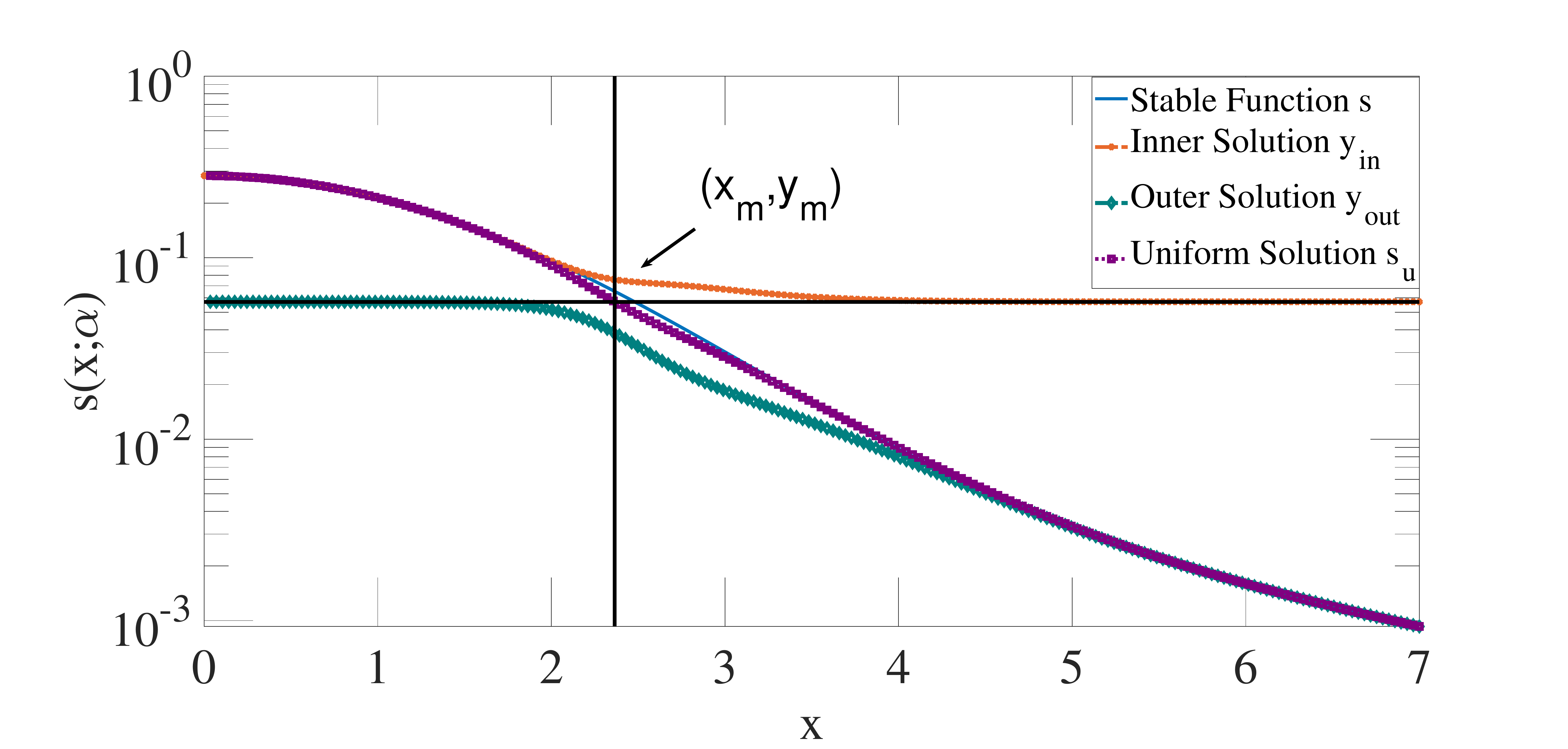}
  \caption{ \textit{{Uniform solution $s_{u}$ for $\alpha=1.80$ as a result of joining the inner solution $y_{in}$ with the outer solution $y_{out}$. The tolerance $\epsilon=10^{-6}$, $\mu=0.4$ and $n=8$.}}}
  \label{fig:Fig05B_Uniform2}
  \end{figure}

The similarity between the exact solution of $s(x;\alpha)$ and the uniform solution $s_{u}(x;\alpha,\epsilon,\mu)$ is observed in Figure \ref{fig:Fig05A_Uniform1} and \ref{fig:Fig05B_Uniform2} for $\alpha=0.75$ and $\alpha=1.80$ respectively. For $\alpha<1$, a good approximation between $s(x;\alpha)$ and $s_{u}(x;\alpha,\epsilon,\mu)$ is obtained in the tails after mixing two different orders. The order for the inner solution is $n_{i}=6$, which makes the solution concave upward. The order for the outer solution is $n_{o}=17$, which makes the solution concave downward.  This combination of orders will ensure a good matching asymptotic behaviour. For $\alpha>1$, the uniform solution works well, and a good uniform solution is obtained quickly with a lower order $n=6$.\par Lower orders can be used for both cases,  where the most important aspect to consider is the different concavity between $y_{in}$ and $y_{out}$ for the matching asymptotic behaviour. The concavity of the inner and outer solution is defined by the trigonometric element in each solution.


  \section{\label{sec:conclusions} CONCLUSIONS}
  In this paper we presented a uniform solution of the \textcolor{black}{L\'evy-stable} distribution. This solution converges to the \textcolor{black}{L\'evy-stable} distribution function in the full range  of $x$ values $-\infty <x <\infty $. This condition makes our uniform solution more robust than previous analytical expressions that were only applicable for extreme values $x\rightarrow0$ or $x\rightarrow\infty$. Also, our uniform solution removes the negative values obtained in previous numerical solutions of the \textcolor{black}{L\'evy-stable} distribution function for all $\alpha$ values, which makes this solution more reliable because a probability density function must be always positive.\par 
The uniform solution is the result of an asymptotic matching between the inner and outer expansions. The inner expansion results from the Taylor series expansion of the characteristic function of the \textcolor{black}{L\'evy-stable} distribution around $x=0$. The outer expansion is obtained from the Taylor expansion of the integrand of the trans-stable function around $t=0$.  The convergence of these expansions is guaranteed if the integrands are truncated, and the speed of convergence depends on how is the truncation implemented.\par
For $\alpha\leq1$, the uniform solution provides a good approximation  for all the range of $x$ values. Also, the numerical integration of the trans-stable function constitutes a second option which allows us to obtain a robust numerical solution of the \textcolor{black}{L\'evy-stable} distribution function and removes the oscillations.
For $\alpha>1$, the uniform solution provides an analytical solution of the \textcolor{black}{L\'evy-stable} distribution function based on fast converging series. \textcolor{black}{ Consequently, the closed-form solution presented in this paper will provide an analytical solution of the fractional kinetic equations (FDE,FDAE,FFPE).}\par
\textcolor{black}{Additionally, having an analytical solution for the L\'evy-stable distribution will contribute on modelling stock markets. To achieve this, L\'evy-stable noise will be generated numerically.  The following procedure is described to generate L\'evy-stable noise. First random points between 0 and 1 are generated. Then, the inverse of the cumulative distribution function (CDF) of the L\'evy-stable distribution is applied to these points. Consequently, the corresponding image of the uniformly generated points will be L\'evy-stable distributed. Different compromises between accuracy and efficiency in the random number generation can be attained by changing the order $n$ in Eq.~(\ref{eq:Su}). Hence, a computational efficiency and high precision are achieved during the generation of large sets of points.} \par \textcolor{black}{For modelling stock markets, L\'evy-stable noise represents the net trading volume ---difference between buy and sell stocks' volume--- and will feed macroscopic models of the stock markets.  
On the other hand, to develop microscopic model of stock markets, the  L\'evy-stable noise can be used to represent the order book (OB) ---list of request for buy and sell orders with prices and volumes---. The use of Levy-Stable noise is justified by the fact that volumes, lifetime of orders,  and the placement of limit orders in a OB present power-law decays with coefficients on L\'evy-stable range. In consequence, a more realistic microscopic model can be developed by the use of our closed-form solution of the L\'evy-stable distribution function. } 
  
\begin{acknowledgments}
F. Alonso-Marroquin thanks Hans J. Herrmann for useful discussions and his hospitality in ETHZ.
\end{acknowledgments}
\bibliography{Natbibk}
\end{document}